\documentclass[fleqn,figuresright]{2023SCGE}
\usepackage{hyperref}%PDF??????
\usepackage{multirow}
\usepackage{supertabular}
\usepackage{amsmath} 
\usepackage{amssymb}
\usepackage{makecell}
\usepackage{soul}
\usepackage{ulem}

\newcommand{\Fermi}{\textit{Fermi}/GBM}
\newcommand{\fermi}{\textit{Fermi}/GBM~}
\newcommand{\T}{$T_{\rm ref}$}

\begin{document}

\ensubject{subject}

%%%%%%%%%%%%%%%%%%%%%%%%%%%%%%%%%%%%%%%%%%%%%%%%%%%%%%%
%%% Authors do not modify the information below
%%% ????????????????
%%% ??????????, ????????????{}, ???????????????????
%Letter to the Editor??Article%??????
\ArticleType{Article}%??Article
\SpecialTopic{SPECIAL TOPIC: }%???????
% \Year{2024}
% \Month{August}
% \Vol{67}
% \No{8}  % 
% \DOI{10.1007/s11433-023-2381-0} % 10.1007/s11433-023-2381-0 
% \ArtNo{289511}  % 289511
% \ReceiveDate{October 27, 2023} % October 27, 2023
% \AcceptDate{March 19, 2024}  % March 19, 2024
% % \OnlineDate{****} # March 20, 2024  

\Year{*}
\Month{*}
\Vol{*}
\No{*}
\DOI{??}
\ArtNo{000000}
\ReceiveDate{****}
\AcceptDate{****}
%%%%%%%%%%%%%%%%%%%%%%%%%%%%%%%%%%%%%%%%%%%%%%%%%%%%%%%

%%% title: ????
%%%   \title{title}{title for citation}
\title{Observation of spectral lines in the exceptional GRB 221009A}{Observation of spectral lines in the exceptional GRB 221009A}
%{Gaussian Emission Line of GRB221009A}

%%% Corresponding author: ???????
%%%   \author[number]{Full name}{{email@xxx.com}}
%%% General author: ???????
%%%   \author[number]{Full name}{}

% \author[1]{A. Author}{{zz@scichina.org}}%
% \author[2]{B. Author}{email@xxx.com}
% \author[1]{C. Author}{}%\protect\\?§Þ??§Ø????
% \author[1]{\\D. Author}{}%
% \author[1]{E. Author}{}

\author[1,2]{Yan-Qiu Zhang}{}%
\author[1]{Shao-Lin Xiong}{{xiongsl@ihep.ac.cn}}
\author[3,4,5]{Ji-Rong Mao}{{jirongmao@ynao.ac.cn}}
\author[1]{Shuang-Nan Zhang}{{zhangsn@ihep.ac.cn}}
\author[1,2]{Wang-Chen Xue}{}%
\author[1,2]{\\Chao Zheng}{}%
\author[1,2]{Jia-Cong Liu}{}% 
\author[1]{Zhen Zhang}{}%
\author[1]{Xi-Lu Wang}{}%
\author[1]{Ming-Yu Ge}{}%
\author[1]{Shu-Xu Yi}{}%
\author[1]{Li-Ming Song}{}%
\author[1]{\\Zheng-Hua An}{}%
\author[6]{Ce Cai}{}%
\author[1]{Xin-Qiao Li}{}%
\author[1]{Wen-Xi Peng}{}%
\author[1,2]{Wen-Jun Tan}{}%
\author[1,2]{Chen-Wei Wang}{}%
\author[1]{\\Xiang-Yang Wen}{}%
\author[1,2]{Yue Wang}{}%
\author[7]{Shuo Xiao}{}%
\author[1]{Fan Zhang}{}%
\author[1,8]{Peng Zhang}{}%
\author[1]{Shi-Jie Zheng}{}%

%%% Author information for page head. ?¨¹?§Ö????????
%%% ??????????????, ??????????author???
\AuthorMark{Y.Q Zhang}%\authorcr????????

%%% Authors for citation. ????????§Ö????????
%%% ??????????????, ??????????author???
\normalem
\AuthorCitation{Y.Q Zhang, S.L Xiong, J.R. Mao, S.N Zhang, et al}

%%% Address. ???
%%%   \address[number]{Address, City {\rm Postcode}, Country}
\address[1]{Key Laboratory of Particle Astrophysics, Institute of High Energy Physics, Chinese Academy of Sciences,  Beijing 100049, China}
\address[2]{University of Chinese Academy of Sciences, Beijing 100049, China}
\address[3]{Yunnan Observatories, Chinese Academy of Sciences, 650011 Kunming, Yunnan Province, China}
\address[4]{Center for Astronomical Mega-Science, Chinese Academy of Sciences, Beijing 100012, China}
\address[5]{Key Laboratory for the Structure and Evolution of Celestial Objects, Chinese Academy of Sciences, Kunming 650216, China}
\address[6]{College of Physics and Hebei Key Laboratory of Photophysics Research and Application, Hebei Normal University, Shijiazhuang, \\ Hebei 050024, China}
\address[7]{Guizhou Provincial Key Laboratory of Radio Astronomy and Data Processing, Guizhou Normal University,\\ Guiyang 550001, China}
\address[8]{College of Electronic and Information Engineering, Tongji University, Shanghai 201804, China}
%\contributions{}%????????

%%% Abstract. ??
\abstract{As the brightest gamma-ray burst ever observed, GRB 221009A provided a precious opportunity to explore spectral line features. In this paper, we performed a comprehensive spectroscopy analysis of GRB 221009A jointly with GECAM-C and \fermi data to search for emission and absorption lines.
For the first time we investigated the line feature throughout this GRB including the most bright part where many instruments suffered problems, and identified prominent emission lines in multiple time intervals. The central energy of the Gaussian emission line evolves from about 37 MeV to 6 MeV, with a nearly constant ratio (about 10\%) between the line width and central energy.  
Particularly, we find that both the central energy and the energy flux of the emission line evolve with time as a power law decay with power law index of -1 and -2 respectively. %, resembling the flux evolution of TeV afterglow in the same time range. 
We suggest that the observed emission lines most likely origin from the blue-shifted %$e^{+}e^{-}$ 
electron positron pair annihilation 511 keV line. We find that a standard high latitude emission scenario cannot fully interpret the observation,
thus we propose that the emission line comes from some dense clumps with electron positron pairs traveling together with the jet.
% Therefore we suggest that the emission lines origin from the $e^{+}e^{-}$ annihilation 511 keV line in some dense clumps probably locating in the external shock region of the jet. 
In this scenario, we can use the emission line to directly, for the first time, measure the bulk Lorentz factor of the jet ($\Gamma$) and reveal its time evolution (i.e. $\Gamma \sim t^{-1}$) during the prompt emission. 
Interestingly, we find that the flux of the annihilation line in the co-moving frame keeps constant. 
These discoveries of the spectral line features shed new and important lights on the physics of GRB and relativistic jet.}   

\keywords{GRB, GRB 221009A, emission line  }

\PACS{98.70.Rz, 95.85.Pw, 52.25.Os}

\maketitle

%\tableofcontents%?????

%%%%%%%%%%%%%%%%%%%%%%%%%%%%%%%%%%%%%%%%%%%%%%%%%%%%%%%
%%% The main text. ???????
%???????????????????\cref{fig1}
%\twocolumn\onecolumn
%%%%%%%%%%%%%%%%%%%%%%%%%%%%%%%%%%%%%%%%%%%%%%%%%%%%%%%

\begin{multicols}{2}
\section{Introduction}\label{sec:Introduction}
% \noindent 
Gamma-ray bursts (GRBs) are astrophysical catastrophic 
events with exceptional power, making them the most energetic explosion known in the universe. 
These events have a typical isotropic luminosity 
ranging from $10^{51}$ to $10^{53}$ erg s$^{-1}$ \cite{Zhangbing_GRB}. 
GRBs can be classified into two types traditionally based on their duration. Those lasting more than 2 seconds (long GRBs) are believed to originate from the dramatic collapse of massive stars \cite{Long_GRB}. Conversely, those with duration shorter than 2 seconds (short GRBs) are thought to arise from the merger of binary neutron stars or neutron star and black hole \cite{Short_GRB}. The study of these different types of GRBs provides valuable insights into the diverse astrophysical processes ending up with a relativistic jet.

GRB 221009A stands out as the most luminous gamma-ray burst ever detected, boasting its bolometric isotropic energies reaching up to a record-breaking $\sim 10^{55}$ erg as accurately measured by GECAM-C and \textit{Insight}-HXMT \cite{HXMT-GECAM:GRB221009A}. The first real-time trigger of GRB 221009A was provided by \textit{Fermi}/GBM, while the real-time trigger of GECAM-C was disabled \cite{HXMT-GECAM:GRB221009A}. The precise localization of this event was first made by the observations of \textit{Swift}/BAT and \textit{Fermi}/LAT, while X-shooter provided accurate redshift measurements \cite{X-shooter}. A multitude of observing facilities, spanning from radio to TeV \cite{2022GCN.32763....1B,2023ApJ...946L..28L,2023ApJ...946L..24W,HXMT-GECAM:GRB221009A,lhaaso2023tera} wavelengths as well as high energy neutrino\cite{2023ApJ...946L..26A}, diligently monitored this extraordinary burst.

However, the exceptional brightness of GRB 221009A posed challenges for many telescopes, leading to various instrumental effects such as pulse pileup, data saturation, etc. Consequently, although there were many observations in X-ray and gamma-ray bands where the GRB radiated its most energy, obtaining precise measurements of the energy of this event proved to be a formidable task. Thanks to the dedicated design of detector and read-out electronics as well as a special operational mode setting, GECAM-C recorded high resolution unsaturated data and provided a uniquely accurate measurement of both the temporal and spectral characteristics \cite{HXMT-GECAM:GRB221009A} of this one-in-thousands-years GRB \cite{2023ApJ...946L..31B}. 

While GECAM-C provided accurate spectral measurement from 15 keV to 5.5 MeV with high-quality data even during the most bright part of GRB 221009A \cite{HXMT-GECAM:GRB221009A}, \fermi could extend the spectral measurements in the higher energy band up to about 35 MeV (with BGO detectors), although a special care should be taken to deal with the GBM spectrum during the bright part of this GRB \cite{Fermi:GRB221009A}. 
To 
\vspace{0.4cm} %
\Authorfootnote
take advantages of better timing and spectral coverage as well 
as cross-checking between instruments, we have performed a joint analysis with GECAM-C and Fermi/GBM data of GRB 221009A, aiming for a more comprehensive spectral and temporal analysis both for the prompt emission and early afterglow phases. Indeed, the GECAM-C and \fermi joint analysis for early afterglow has revealed unprecedented details of the temporal and spectral properties of the early afterglow \cite{Afterglow_zhengchao}, including a more accurate measurement of the jet break time (i.e. 1246$^{+27}_{-26}$ s after the \textit{Insight}-HXMT trigger time\footnote{\textit{Insight}-HXMT trigger time is 2022-10-09T13:17:00.050 UTC (denoted as $\rm T_{tri}$).}).

Spectral lines are of critical importance in revealing the physics of GRBs. Searching for spectral lines have been intensively studied, however, neither emission line nor absorption line has been confirmed before the observation of GRB 221009A \cite{2004AdSpR..34.2696B,1999ASPC..190..133B}. With the record-breaking brightness \cite{HXMT-GECAM:GRB221009A}, GRB 221009A offers a unique opportunity to accurately characterize the spectrum and identify line features. Among our joint analysis of GECAM-C and \Fermi, search for spectral lines throughout the burst has been given a special attention. While we found some interesting excess features in the spectra yet to be carefully identified, Ravasio et al. reported the discovery of about 10 MeV lines based on \fermi data alone \cite{edvige2023bright}. However, their study was limited to the less-bright part of GRB 221009A, because \fermi data suffered instrumental effects during the bright episode, preventing them from exploring the line features during the most bright and interesting part of this GRB.

Here we report a comprehensive analysis of spectral lines of GRB 221009A jointly with GECAM-C and \fermi. We applied the accurate measurement of GECAM-C to correct the \fermi instrumental effects during the bright part of GRB 221009A, which enabled us to explore the complete evolution of the spectral lines throughout the full course of the prompt emission with the time range of \T + (0, 600) s, where $T_{\rm ref}$ is set to 2022-10-09T13:17:00.000 (UTC) for convenience\footnote{Please note that this reference time (\T) is chosen to be a rounded second to align with the time range of binned data, thus is slightly different from the \fermi trigger time (2022-10-09T13:16:59.990 UTC) or \textit{Insight}-HXMT trigger time (2022-10-09T13:17:00.050 UTC).}. 

We find that significant emission lines could be detected at three time intervals, namely G1 for \T+(246, 256) s,  G2 for \T+(270, 275) s, and  G3 for \T+(275, 360) s. Importantly, we reveal a remarkable time-evolution of the line energy which could be described with a powerlaw function. 

This paper is organized as follows: In \cref{sec:Observation} we first describe the observations of GRB221009A by GECAM-C and \fermi. Details of the data analysis, including time intervals, background and spectral fitting process are discussed in \cref{sec:Data Analysis}. The results of the spectral line analyses are presented in \cref{sec:results}. Discussion and conclusion are given in \cref{sec:Discussion ans Conclusions}, followed by a summary in \cref{sec:Summary}.

\vspace{-0.5cm} %
\section{Observations}\label{sec:Observation}
\vspace{-0.2cm} %
\subsection{GECAM-C Observation}\label{sec:GECAM-C}
 GECAM (Gravitational wave high-energy Electromagnetic Counterpart All-sky Monitor) is a constellation of gamma-ray monitors funded by the Chinese Academy of Sciences.
Its primary scientific objective is to systematically monitor and study a diverse range of high energy transients, including Gamma-Ray Bursts (GRBs, e.g.\cite{HXMT-GECAM:GRB221009A}), Soft Gamma-ray Repeaters (SGRs, e.g.\cite{Minimum_Variation_Timescales_xiao}), Solar Flares (SFLs, e.g.\cite{quasi-periodic_SF_zhaohs}), X-ray Binaries (XRBs, e.g.\cite{XRBs}), Terrestrial Gamma-Ray Flashes and Terrestrial Electron Beams (TGFs/TEBs, e.g.\cite{TGF_TEB_zhaoyi}), etc.
 The first two micro-satellites, GECAM-A and GECAM-B, were launched on December 10, 2020 \cite{li2022technology}. As the third instrument of GECAM constellation, GECAM-C (also called HEBS) \cite{zhangperformance} was launched onboard the New Space Technology Experimental Satellite (SATech-01) on July 27, 2022 \cite{SATech-01}. GECAM-C is composed of two detector domes configured with a total of 12 GRDs \cite{an2022design} (Gamma-ray detectors, made with LaBr$_3$ or NaI scintillators readout by SiPM array) and 2 CPDs \cite{xu2022design} (Charge particle detectors, made with plastic scintillators readout by SiPM array). It is worth noting that 10 GRDs of GECAM-C have two electronic readout channels with different detection energy range, which are often referred to as high-gain (HG) and low-gain (LG) respectively. In addition, a detailed calibration for GECAM-C has been performed, which shows that the GECAM-C has good temporal and spectral performance \cite{zheng:2023ground_calibration,Zhangyq:2023cross_calibration,2023arXiv230811362X}.

Thanks to the dedicated designs of detector and electronics as well as a special working mode for the high latitude region\footnote{Only one GRD (i.e. GRD01) and one CPD (i.e. CPD02) are turned on to collect data normally when the satellite passes through this orbital region.}, GECAM-C accurately measured the GRB 221009A and found it is the brightest gamma-ray burst ever recorded \cite{GCN.32751,HXMT-GECAM:GRB221009A}. 
%\footnote{\url{https://gcn.gsfc.nasa.gov/gcn3/32751.gcn3}}. 
The low gain channel of the GRD01 is not saturated, the dead time recording of low gain is accurate, and the pulse pileup effect of low gain is also negligible. There is an issue of incorrect recording of deadtime for the high gain readout, resulting in a slightly high count rate during the bright part of the burst. However, we can correct high gain data with the accurate low gain data. Detailed analysis processes could be found in a previous work \cite{HXMT-GECAM:GRB221009A}.

\vspace{-0.2cm} %
\subsection{\fermi Observation}\label{sec:Fermi}

Fermi Gamma-ray Burst Monitor (\textit{Fermi}/GBM) 
is one of the two major instruments of the Fermi Gamma-ray Space Telescope. \Fermi \,has a very wide band measurement capability, with 12 NaI(Tl) detectors (labeled as n0, n1,..., na, nb) covering an energy range of 8-1000 keV and 2 BGO detectors (labeled as b0,b1) with the energy range of ~0.2-40 MeV \cite{meegan2009fermi}. 

\Fermi \,was triggered by GRB 221009A at 2022-10-09T13:19:59.990 (UTC)\cite{GCN.32636,GCN.32642}.
%\footnote{\url{https://gcn.gsfc.nasa.gov/gcn3/32636.gcn3}},  %\footnote{\url{https://gcn.gsfc.nasa.gov/gcn3/32642.gcn3}}). 
Unfortunately, it has very severe saturation data loss and pulse pileup effects during the bright part of GRB 221009A, including the main burst and peak region of the flare. The \fermi team has officially released some caveats and bad time interval (BTI) for the GBM data analysis of  GRB 221009A\footnote{\url{https://fermi.gsfc.nasa.gov/ssc/data/analysis/grb221009a.html}\label{fn:GBM}}. They also performed a detailed analysis of the GBM data and tried to repair their data during BTI, but their corrected light curves (Fig 7 in \cite{Fermi:GRB221009A}) of the two main peaks are apparently inconsistent with that of GECAM-C \cite{HXMT-GECAM:GRB221009A} and other monitors (e.g. Konus-Wind \cite{2023ApJ...949L...7F,2023GCN.33427....1S}). Here we use the GECAM-C accurate data to calibrate the GBM data, resulting in a reasonable and consistent joint spectroscopy analysis.

\vspace{-0.5cm} %
\section{Data Analysis}\label{sec:Data Analysis} 

\vspace{-0.2cm} %
\subsection{GTI and BTI}\label{sec:GTI}
\begin{figure}[H]%h-t-b-p
\centering
\includegraphics[width=0.85\columnwidth]{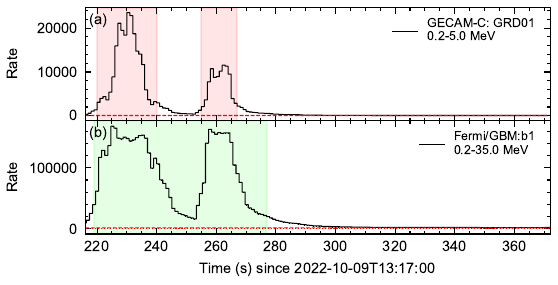}%"scale"width=1.0\columnwidth"
\caption{GECAM-C and \fermi raw light curves of GRB 221009A. Panel (a) shows the light curve of the GECAM-C GRD01 detector. The black line is the total light curve in low gain, while the red dashed line is the background. The pink shaded area is the time period where the high-gain dead time recording is incorrect while the low gain data is free of this issue. Panel (b) is the light curve of the \fermi b1 detector, which suffers significant data saturation during the bright epochs. The green shaded area is the BTI provided in the \fermi official website\footref{fn:GBM}.}.
\label{fig:GTI}%{}"fig:example1"\ref{fig:example1}
\end{figure}

GTI (good time interval) means a period of time when the observation data is good for analysis, while BTI (bad time interval) is the opposite of GTI. These GTI and BTI of each instrument should be considered carefully when conducting data analysis of this extremely bright GRB 221009A. The detailed information is displayed in \cref{fig:GTI} and \cref{Detailed Information}.

\begin{table*}[t]
\begin{center} 
\footnotesize    
\begin{threeparttable}\caption{Detailed Data Analysis Information}
\label{Detailed Information}
\tabcolsep 45
pt %space between two columns. ????????§Þ???
% \doublerulesep 0.1pt \tabcolsep 13pt
\begin{tabular*}{\textwidth}{ccc}
\toprule
   & GECAM-C & \Fermi \\
   \hline 
Dectectors & GRD01 & b0,b1 \\
Energy Range\tnote{2)} & \makecell[c]{15-300 keV (high-gain) \\ 0.7-5.5 MeV (low-gain)} & 4-35 MeV   \\
BTI & - \tnote{1)} & (219, 277) s \\
Software & \texttt{GECAMTools} & \texttt{GBM Data Tools} \cite{GBMDATATools} \\
Data file type \tnote{2)} & Bspec & TTE/Cspec \\
\bottomrule
\end{tabular*}
\begin{tablenotes}
\item[1)] For GECAM-C, there is only a problem with high-gain dead time recording in the \T+(220, 240) s and \T+(255, 267) s time periods, but this does not affect the data analysis.
\item[2)] Various energy ranges and data file types are employed for various time periods, as stated in the Appendix.

\end{tablenotes}
\end{threeparttable}
\end{center}
\end{table*}

\begin{table*}[t]
\begin{center} 
\footnotesize    
\begin{threeparttable}\caption{Detailed Data Analysis Information for Spectral Fitting}
\label{Detailed Spectrum Information}
% \tabcolsep 11pt %space between two columns. ????????§Þ???
% \doublerulesep 0.1pt \tabcolsep 13pt
\begin{tabular*}{\hsize}{@{}@{\extracolsep{\fill}}cccccccc@{}}
\toprule
  \multirow{2}{*}{Time Range} & \multirow{2}{*}{Model} &\multicolumn{3}{c}{GECAM-C}  &\multicolumn{3}{c}{\Fermi}\\
  \cline{3-5} \cline{6-8}
  & & Det & Data Type & Energy Range & Det & Data Type & Energy Range\\
   \hline 
(246,256) s & band+gauss & GRD01 & bspec & HG: 15-300 keV, LG: 0.7-5.5 MeV \tnote{1)} & b0,b1 & Cspec & 4-35 MeV\\
(270,275) s & band*gabs+gauss & GRD01 & bspec &  HG: 150-300 keV, LG: 0.7-5.5 MeV & b0,b1 & Cspec & 4-35 MeV\\
(275,280) s & band+gauss & GRD01 & bspec &  HG: 150-300 keV, LG: 0.7-5.5 MeV & b0,b1 & TTE & 4-35 MeV\\
(280,285) s & band+gauss & GRD01 & bspec &  HG: 150-300 keV, LG: 0.7-5.5 MeV & b0,b1 & TTE & 4-35 MeV\\
(285,290) s & band+gauss & GRD01 & bspec & HG: 15-300 keV, LG: 0.7-5.5 MeV & b0,b1 & TTE & 2-35 MeV\\
(290,295) s & band+gauss & GRD01 & bspec & HG: 15-300 keV, LG: 0.7-5.5 MeV & b0,b1 & TTE & 1-35 MeV\\
(295,300) s & band+gauss & GRD01 & bspec & HG: 15-300 keV, LG: 0.7-5.5 MeV & b0,b1 & TTE & 1-35 MeV\\
(280,300) s & band+gauss & GRD01 & bspec & HG: 15-300 keV, LG: 0.7-5.5 MeV & b0,b1 & TTE & 4-35 MeV\\
(300,310) s & pl+gauss & GRD01 & bspec & HG: 15-300 keV, LG: 0.7-5.5 MeV & b0,b1 & TTE & 1-35 MeV\\
(310,320) s & pl+gauss & GRD01 & bspec & HG: 15-300 keV, LG: 0.7-5.5 MeV & b0,b1 & TTE & 1-35 MeV\\
(300,320) s & pl+gauss & GRD01 & bspec & HG: 15-300 keV, LG: 0.7-5.5 MeV & b0,b1 & TTE & 1-35 MeV\\
(320,340) s & cutoffpl+pl+gauss & GRD01 & bspec & HG: 15-300 keV, LG: 0.7-5.5 MeV & b0,b1 & TTE & 1-35 MeV\\
(340,360) s & cutoffpl+pl+gauss & GRD01 & bspec & HG: 15-300 keV, LG: 0.7-5.5 MeV & b0,b1 & TTE & 1-35 MeV\\
(320,360) s & cutoffpl+pl+gauss & GRD01 & bspec & HG: 15-300 keV, LG: 0.7-5.5 MeV & b0,b1 & TTE & 1-35 MeV\\
\bottomrule
\end{tabular*}
\begin{tablenotes}
\item[1)] HG: high-gain, LG: low-gain; 
\end{tablenotes}
\end{threeparttable}
\end{center}
\end{table*}
\vspace{-0.4cm}

\vspace{-0.4cm}
For GECAM-C, the GTI covers the whole burst. Although there are some problems with the dead time recording of the GRD01 high-gain readout during \T+(220, 240) s and \T+(255, 267) s, as shown in the pink shaded area of \cref{fig:GTI}, we can correct the high-gain dead time effect using the accurate low-gain data\cite{HXMT-GECAM:GRB221009A}. 

\vspace{-0.2cm} %
\begin{figure}[H]%h-t-b-p
\centering
\includegraphics[width=0.84\columnwidth]{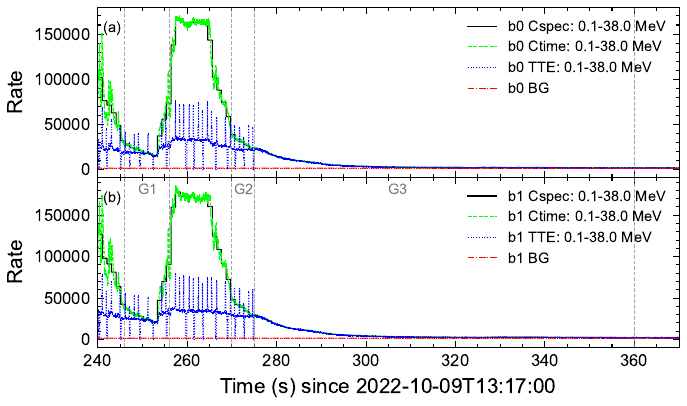}%"scale"width=1.0\columnwidth"
\caption{Light curves of the \fermi BGO detectors (b0 and b1) for the \T+(240, 370) s time period. The G1, G2 and G3 denote the time periods of \T+(246, 256) s, \T+(270, 275) s, and \T+(275, 360) s, respectively. The black, green and blue lines represent the Cspec, Ctime and TTE data types of \Fermi. Panels (a) and (b) show the light curves of b0 and b1, respectively. Notably, we can see from the light curves that the \fermi TTE data has a much more severe data loss than Cspec and Ctime data until \T+275 s. After \T+275 s these three types of GBM data show good agreement.} 
\label{fig:LC_For_GBM}%{}"fig:example1"\ref{fig:example1}
\end{figure}

For \fermi data, the officially declared BTI is \T+(219,277) s, where significant saturation and pipe-up effects occurred. In our analyses, we find that these effects do not affect the spectral shape above a certain energy when the energy spectrum shape changes slowly. We find that, when jointly fitting the \fermi and GECAM-C spectrum, the constant factor between these two instruments remains approximately unity for the time intervals where \fermi is free of data issues, which demonstrates good consistency between \fermi and GECAM-C and also validates our data analysis procedure.

As displayed in \cref{fig:LC_For_GBM}, we present the light curves for the three types of data from \fermi. It shows that there is good consistency between the GBM Cspec and Ctime data, however they both suffered some data loss around the peak region when compared to the light curves of GECAM-C in \cref{fig:GTI}. On the other hand, the GBM TTE data experienced a much more significant data loss before \T+275 s. Additionally, it is important to note that while the \fermi Cspec trigger data \cite{meegan2009fermi} boasts a temporal accuracy of 1.024 s, it is constrained by a fixed temporal boundary, making it less flexible to choose time range of data. Therefore, we utilize the Cspec data for the time period when the TTE data is severely compromised before \T+275 s, and use the TTE data after \T+275 s to improve the time accuracy.

\subsection{Background}\label{sec:BG}
Since the space particle environment in the high latitude region is very complicated, the background evolution of gamma-ray detectors may be so irregular that it is very difficult to estimate background during the burst by fitting the background with an empirical polynomial function, especially for a long duration burst. Fortunately, the SATech-01 satellite was making a pointed observation at the time of GRB 221009A detection, thus GECAM-C remained almost a constant pointing direction during the entire course of GRB 221009A as well as the revisit orbits (see more details in \cite{Afterglow_zhengchao}). This allows us to use the detector count rate of the revisit orbits to estimate the background during the GRB 221009A. This approach has been proven to be able to provide a good description of the background during the burst in our previous work \cite{HXMT-GECAM:GRB221009A,Afterglow_zhengchao}.

\begin{figure*}[t] % [H]%h-t-b-p
\centering
\includegraphics[width=1.9\columnwidth]{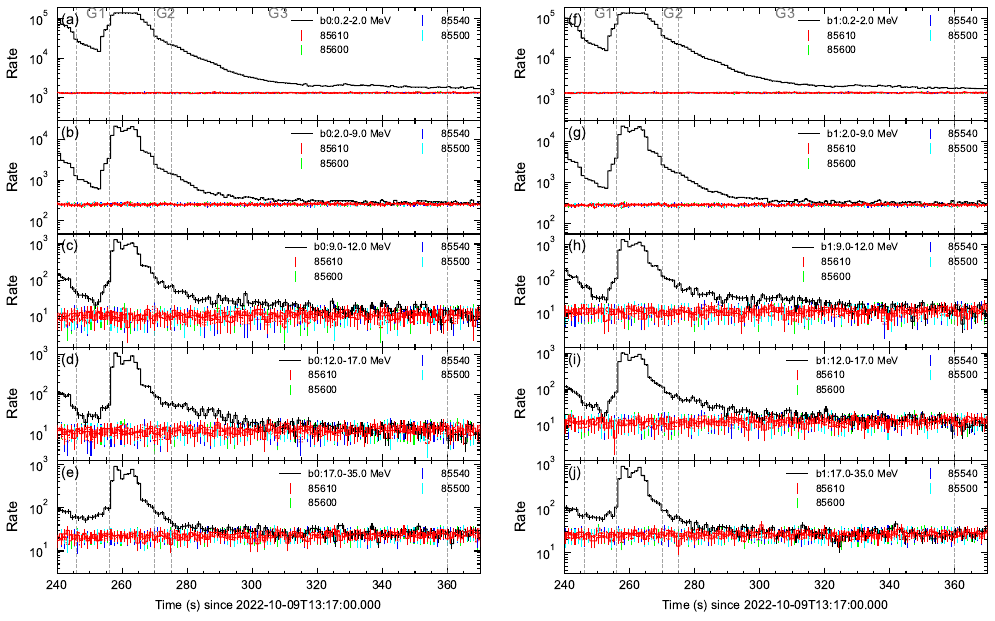}%"scale"width=1.0\columnwidth"
\caption{Multi-energy segment light curves of \fermi with b0 and b1 detectors in the time period \T+(240, 370) s. The definition of the G1, G2 and G3 are the same as \cref{fig:LC_For_GBM}. Panels (a), (b), (c), (d) and (e) are light curves of b0 and panels (f), (g), (h), (i) and (j) are light curves of b1. The red, green, blue, and cyan colors represent the four revisit times: 85610~s, 85600~s, 85540~s and 85500~s, respectively. The data in red are the actual used background data for each energy band: they are averaged from the data of the revisited orbits before and after the current one.} 
\label{fig:LC_For_ME_GL_GBM}%{}"fig:example1"\ref{fig:example1}
\end{figure*}

For the background of \Fermi, we conducted a thorough analysis and reached two conclusions \cite{Afterglow_zhengchao}: First, estimating the background by averaging the two neighboring revisit orbits is recommended.
%as relying solely on either the front or back would introduce bias. 
Second, we tried four revisit times and found that, upon averaging, the background values at these times remained consistent within the margin of error. In our final data analysis we used 85610 s as the revisit time. Details are displayed in \cref{fig:LC_For_ME_GL_GBM}. 

We noticed that the multi-energy band light curve of the G1 interval is interesting, as shown in \cref{fig:LC_For_ME_GL_GBM}.
%, and the rest will be explained in detail later. 
There is an additional excess on the evolutionary trend in panels (e) and (j).  For most energy bands the net counts (total light curve counts minus background counts) decrease as energy increases, but there is an excess in the net counts above 17 MeV. 
We notice that this excess is not only relative to the energy evolutionary (light curves in different energy band), but also relative to the time evolution of the counts prior to this time period (before \T+246 s). 
Therefore, this excess should not be caused by high count rate effect, otherwise it should also appear a few seconds before this time period where the count rate is even higher.
Indeed, we find that this excess corresponds to an emission line component, as shown below. 

\subsection{Spectral Fitting}\label{sec:Spectrum Fitting}

\subsubsection{Spectral Models}\label{sec:Model}

In our spectral analysis, Band model \cite{band1993batse}, cutoff power-law model (denoted as \textit{cpl}) \cite{von2014second}, power-law model (denoted as \textit{pl}), gauss model (for emission line) and gabs model (for absorption line) are employed for the spectral fitting. The analysis tool is \texttt{pyxspec}\footnote{\url{https://heasarc.gsfc.nasa.gov/xanadu/xspec/python/html/index.html}} based on \texttt{\textit{python}}, and the environment is \texttt{xspec} v12.12.0 \cite{1996Xspec} . The spectral models mentioned above are shown in \cref{equ:band_Model,equ:Cutoffpl_Model,equ:pl_Model,equ:gauss_Model,equ:gabs_Model}:
\begin{small}
\begin{equation}
    N_{\rm band}(E)= \begin{cases} 
    A \bigg (\frac{E}{E_{\rm piv}}\bigg )^{\alpha} \exp\bigg(-\frac{E}{E_{\rm c}}\bigg ), & (\alpha -\beta)E_{\rm c} \geq E, \\ 
    A \bigg [\frac{(\alpha-\beta) E_{\rm c}} {E_{\rm piv}}\bigg ]^{\alpha-\beta} \exp(\beta-\alpha)\bigg (\frac{E}{E_{\rm piv}}\bigg )^{\beta}, & (\alpha -\beta)E_{\rm c} \leq E, \\
    E_{\rm piv} = 100 \enspace {\rm keV},
    \end{cases},
    \label{equ:band_Model}
\end{equation}
\end{small}
where $A$ is the normalization amplitude constant ($\rm photons \cdot cm^{-2} \cdot s^{-1} \cdot keV^{-1}$), $\alpha$ is the \textit{low-energy} power-law index and $\beta$ is the \textit{high-energy} power-law index, $E_{\rm c}$ is the characteristic energy in keV, $E_{\rm piv}$ is the pivot energy in keV and usually the value is taken as 100 keV.
\begin{equation}   % cutoffpl Model
N_{\rm cpl}(E)=A(\frac{E}{E_{0}})^{\alpha} \exp(-\frac{E}{E_{\rm c}}),
\label{equ:Cutoffpl_Model}
\end{equation}
where $E_{0}=1$~keV, $A$ is the normalization amplitude constant ($\rm photons \cdot cm^{-2} \cdot s^{-1} \cdot keV^{-1}$) at 1 keV, $\alpha$ is the power law photon index, and $E_{\rm c}$ is the characteristic energy of exponential roll-off in keV. 
\begin{equation}  % pl Model
N_{\rm pl}(E)=A(\frac{E}{E_{0}})^{\alpha},
\label{equ:pl_Model}
\end{equation}
where $E_{0}=1$~keV, $A$ is the normalization amplitude constant ($\rm photons \cdot cm^{-2} \cdot s^{-1} \cdot keV^{-1}$) at 1 keV, $\alpha$ is the photon index of power law.
%(which is a dimensionless quantity).
\begin{equation}  % gauss Model
N_{\rm gauss}(E)=A\frac{1}{\sigma\sqrt{2\pi}}\exp(-\frac{(E-E_{\rm line})^2}{2\sigma^2}),
\label{equ:gauss_Model}
\end{equation}
where $A$ is the flux ($\rm photons \cdot cm^{-2} \cdot s^{-1}$) in the line, $E_{\rm line}$ is the line energy in keV and $\sigma$ is the line width in keV.
\begin{equation}  % gabs Model
N_{\rm gabs}(E)=\exp(-\frac{E_{\rm depth}}{\sqrt{2\pi}\sigma}\exp(-\frac{(E-E_{\rm line})^2}{2\sigma^2})),
\label{equ:gabs_Model}
\end{equation}
where $E_{\rm depth}$ is line depth in keV, $E_{\rm line}$ is the line energy in keV and $\sigma$ is the line width in keV.

\subsubsection{Spectral Fitting Strategy}\label{sec:Fitting}
 In our analysis, we performed a search for Gaussian emission and absorption lines throughout the prompt emission of GRB221009A, including the precursor, main emission and flaring emission. We find that gaussian emission lines only appear in the following time periods: \T+(246, 256) s and \T+(270, 360) s. In the fitting process, we used and compared two groups of spectral models: a single continuum spectrum without line and a continuum spectrum plus a gaussian line component. 

Note that we choose different continuum spectral models depending on the time period of this burst to account for the spectral evolution. For the time interval \T+(246, 256) s, we use the the model \textit{band+gauss}, where the \textit{band} is used to fit the continuum spectrum and the \textit{gauss} is used to fit the emission lines. 
In addition, for the time period \T+(270, 300) s, when the afterglow is less intense and the prompt emission dominates, we also use the \textit{band+gauss} model to fit the continuum and gaussian emission lines separately. For the time period \T+(300, 320) s, the \textit{pl+gauss} model is used. Finally, for the time period \T+(320, 360) s when the prompt emission is not so strong, the \textit{cpl+pl+gauss} model is used for spectral fitting, where the additional \textit{pl} model is invoked to describe the afterglow component. Details are shown in \cref{Detailed Spectrum Information}.

It is important to note that, in order to correct \fermi data we liberalize a constant factor between GECAM-C and \fermi spectra. For \fermi background estimation, we also applied the revisit orbit method, which has been verified \cite{Afterglow_zhengchao}. Moreover, we used two GBM BGO detectors in the fit because both detectors have similar incidence angles for this burst. Since there is inconsistency between two BGOs in lower energies (see \cref{fig:Specfit-examples}), BGO data below a certain energy (refer to the Energy Range column of \cref{Detailed Spectrum Information}) is ignored in our spectral fit, and the energy threshold is determined from the consistency check between the two detectors (b0 and b1) and GECAM-C (see \cref{fig:Specfit-examples}). As for GECAM-C, the available energy range is 15-300 keV for high-gain and 0.7-5.5 MeV for low-gain. Note that 37-40 keV is ignored for all time intervals due to the absorption edge of LaBr$_3$ crystal of GECAM-C GRD detector.

\vspace{-0.4cm}
\section{Results}\label{sec:results}

\subsection{Identification of Spectral Lines}
After checking the time-resolved spectra throughout the burst, we find that the spectral lines could be detected in these time ranges: 246-256 s (G1), 270-275 s (G2) and 275-360 s (G3), and the central energy of the emission line evolves from about 37 MeV to about 6 MeV.

The detailed spectra analysis results for these time intervals are summarized in \cref{Spectrum fitting results with Gauss Component} and \cref{Spectrum fitting results without Gauss Component}, where the results of fitting with the continuum spectrum plus Gaussian emission lines, and with the continuum spectrum only are presented respectively.
We calculated the chance probability value ($p$-value) of the spectral line components through the simulation of likelihood ratio; the results are displayed in \cref{Spectrum fitting results with Gauss Component} and the detailed calculation process can be found in \cref{sec:Sigcal}. The results indicate moderate to high significance of the emission line in most of the time intervals. We emphasize that, regardless of the significance in individual time interval, the reality of these emission lines is very solid as we find a very regular time-evolution of the line central energy and flux shown below. None of any known instrumental effects or background fluctuation can explain such behavior.
% we also calculated the $\rm \Delta AIC $ = $\rm AIC_{Non-gauss}$ - $\rm AIC_{gauss}$ \cite{liddle2007information}.

It is worth to point out that, the energy resolution (FWHM is about 2.36$\sigma$ for gaussian peak) of the BGO near MeV range is about 10\% \cite{2009ExA....24...47B}, which is much less than the detected spectral line widths. In addition, the spectral fitting already considered the detector energy resolution. Thus the line width obtained here is the intrinsic width of the spectral line itself. 

Besides, we also carefully checked for absorption line and do not found any significant absorption line except for a marginal detection in a short time interval (see \cref{sec:GECAM-C check} for details). We note that this is the only absorption line feature we found throughout this burst, however, the very low significance given by the present analysis prevents us from drawing any conclusion. Therefore we will not discuss this absorption feature in the rest of this paper.

\begin{figure}[H]%h-t-b-p
\centering
\includegraphics[width=0.9\columnwidth]{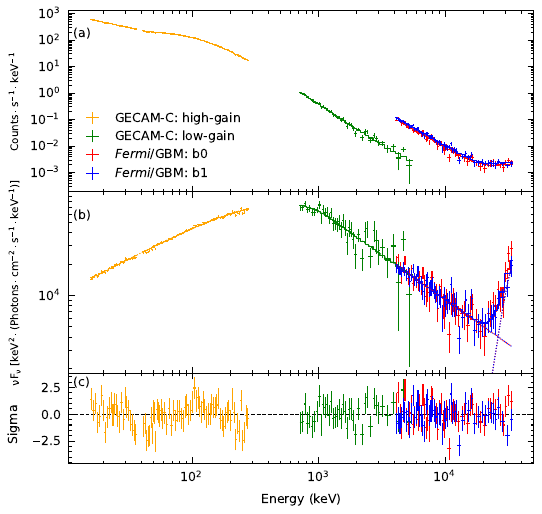}%"scale"width=1.0\columnwidth"
\caption{The spectral fitting result with \textit{band+gauss} spectrum model within the time interval \T+(246, 256) s. The orange and green data points are the high and low gain of the GECAM-C, respectively. The red and blue data points are the two BGO detectors. The solid lines correspond to the model value of each instrument. Panel (c) shows the residuals in terms of $\sigma$ with $1\sigma$ error bars. The residual fluctuates around zero, which means the model fits the data points very well. As shown in this figure, there is one component above 17 MeV that exceeds the expected value.}
\label{fig:32MeV}%{}"fig:example1"\ref{fig:example1}
\end{figure}

% \vspace{-0.7cm}  
\subsection{Discovery of Emission Line during the main burst}\label{sec:Emission in ME}

Whether there is spectral line during the main burst of GRB 221009A is an important question, but searching for it is a very challenging task because many instruments (including \Fermi) suffered data problems during the bright part of the burst. Since GECAM-C provided accurate measurement of the main burst of GRB 221009A \cite{HXMT-GECAM:GRB221009A}, we could, for the first time, search for spectral line features during the most luminous part of this GRB jointly with GECAM-C and \fermi data.
As shown in \cref{fig:LC_For_ME_GL_GBM}, significant excess is found in the light curve (246-256 s) during the main burst as mentioned above. This is a hint of the existence of a new spectral component, such as afterglow or emission line. 

\begin{figure*}[t]
\centering
\includegraphics[width=1.4\columnwidth]{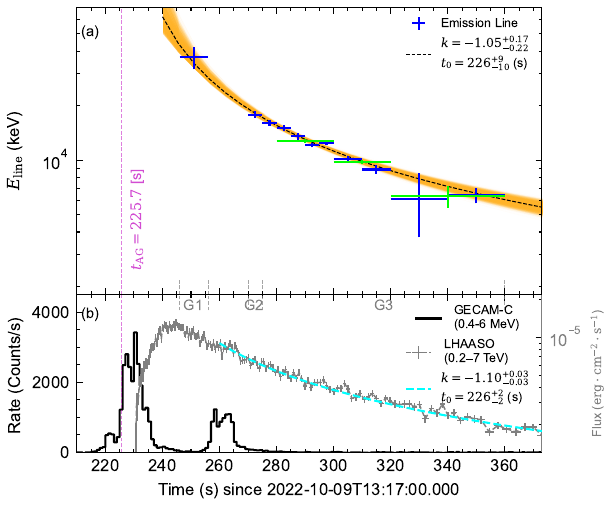}
\caption{Panel (a): Evolution of central energy of emission lines. The blue crosses are the central energy of emission lines at different time periods with $1\sigma$ error bars, the green crosses are the central energy after some time integration for the corresponding time periods with $1\sigma$ error bars, the black lines are the power-law decay fitted model, the orange area shows the $1 \sigma$ confidence interval of the fit model. Panel (b) shows the light curves of GECAM-C (black line) and LHAASO (gray line), and displayed with the corresponding color axes. The vertical purple dash line is the afterglow onset time ($t_{\rm AG}$) determined by TeV afterglow analysis \cite{lhaaso2023tera}.} 
\label{fig:Gauss_Evalution}
\end{figure*}

We perform spectral analysis and also find a large excess structures above 20 MeV in both BGO detectors of \Fermi, as shown in \cref{fig:32MeV}.
Based on the TeV afterglow observation \cite{lhaaso2023tera}, the afterglow reaches its peak at about \T+244 s in the light curve and then decay (see \cref{fig:Gauss_Evalution}). %Therefore, in order to exclude the afterglow as a factor, 
We tried with an additional \textit{pl} model to fit two BGOs data of \fermi above 20 MeV, resulting in a power-law index of $-0.08 \pm 0.45$,
which significantly deviates from the afterglow spectrum (power-law index of afterglow is about -1.7 in MeV band \cite{Afterglow_zhengchao}). These results exclude the afterglow as the origin of this excess.

 Then we testify whether the excess could be explained by emission line. Apart from the continuum component describing the prompt emission, we used a \textit{gauss} emission line component to fit the excess above 20 MeV. The best values of the  central energy and width of this emission line are $37.2 \pm 5.3$  MeV and $7.1 \pm 3.1$ MeV, respectively.
We note that, although this Gaussian emission line does not show its complete shape in the measured energy range of BGO detector, a relatively large part of the left side of Gaussian shape is clearly seen which could be used to constrain parameters of this emission line. As discussed later, the parameters of this emission line (e.g. central energy, line flux, the ratio between line width and central energy) follow the evolution trend of emission lines in other time intervals very well, lending further support to the reality and measurement of this line.

% \vspace{-0.4cm}

\subsection{Evolution of Emission Line Central Energy and Flux}

Thanks to the wide time coverage of the emission line, especially the detection of emission line in the main burst region (see \cref{sec:Emission in ME}), we notice that the emission lines follow a very regular and interesting time evolution.
First, we plot the spectral energy distribution (SED) $\nu F_{\nu}$ for different time intervals with spectral line, as shown in \cref{fig:Gauss_Evalution_vfv}. One can clearly see the evolution of the emission lines as well as continuum component in different time intervals. The time evolution of the parameters of line and continuum components are shown in \cref{fig:Gauss_Evalution} and \cref{fig:Gauss_Evalution_flux}. 
Remarkably we find that both the central energy and the flux of emission lines follow a power law decay with time, which could be fit with \cref{equ:exp_model}:

\begin{figure}[H]
\centering
\includegraphics[width=1\columnwidth]{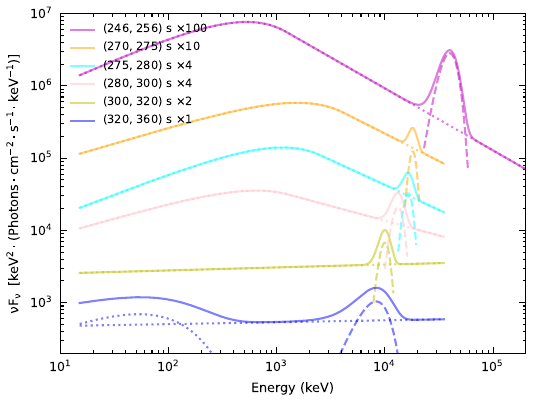}
\caption{$\nu F_{\nu}$ for different time intervals. Different fitting models are used for different time periods. All dashed lines indicate detailed model components, where the dashed line is the line component and the dotted dashed line is continuum spectral component. For \T+(280, 300) s, we only draw a total $\nu F_{\nu}$ instead of five small time intervals for clarity. To better show the spectra, the $\rm \nu F_\nu$ for various time periods are multiplied by different factors shown in the legend.}
\label{fig:Gauss_Evalution_vfv}
\end{figure}

% \begin{scriptsize}
\begin{equation}   
f(t)=10^m\cdot (t-t_0)^{k},
\label{equ:exp_model}
\end{equation}
% \end{scriptsize}
where $m$ is the normalization factor, $k$ is the power law index, and $t_0$ is the initial time of this decay.

We use the \texttt{emcee v3.1.1} \cite{emcee} package to fit the power law decay with the \texttt{MCMC} method. For the emission line central energy evolution, we get the power law index $ k = {-1.05}^{+0.16}_{-0.22}$ and decay initial time $ t_0 = {226}^{+8}_{-10}$~s. For the emission line flux decay, the power law index $ k = {-2.16}^{+0.09}_{-0.12}$ when we fixed the initial time $ t_0 = 226$~s because the error of line flux is relatively larger than the central energy. These fit results are shown in panel (a) of \cref{fig:Gauss_Evalution} and panel (a) and (d) of \cref{fig:Gauss_Evalution_flux}.
%and the orange shaded area is the confidence interval of the fit. 
The MC posterior probabilities of the fitted parameters (\cref{fig:corner}) show that all parameters are well constrained. 

It is particularly interesting that the emission line energy power law decay parameters (both the power law index $k$ and the initial time $t_0$) are in excellent agreement with the power law decay of the TeV afterglow in the same time range as the emission line. Note that we fit the TeV flux evolution with the same function \cref{equ:exp_model} (shown in panel (b) of \cref{fig:Gauss_Evalution}, see \cref{sec:LHAASO Flux Fit} for details), and our fitting results (including the power law decay index and the initial time) of TeV afterglow are well consistent with the results reported by LHAASO team \cite{lhaaso2023tera}, validating our analysis results.

\begin{figure}[H]
\centering
\includegraphics[width=0.91\columnwidth]{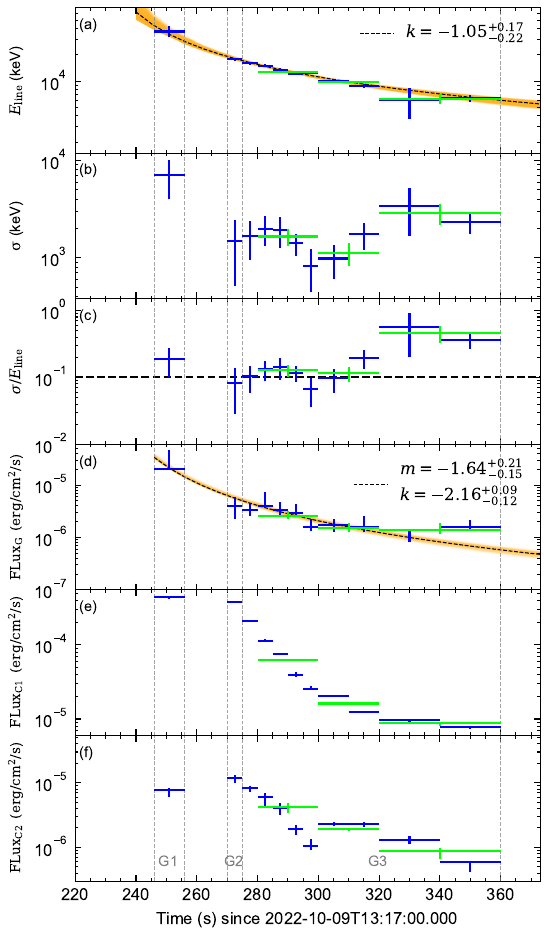}
\caption{Evolution of parameters of the emission line and continuum component. The data points and data lines are defined as in \cref{fig:Gauss_Evalution}. Panels (a) and (b) depict the evolution of the central energy and $\sigma$ of the emission line over time, respectively. Panel (c) displays the ratio of $\sigma$ and line energy, with the dashed line representing 10\%. In the fitted panel (d), we fix $t_0$ to be the result in the $E{\rm line}$ fit. $\rm FLux_{G}$ is calculated in the gaussian 3$\rm \sigma$ energy range; due to the incompleteness of gaussian emission line detection, it is normal that the upper error range of the measurement will be somewhat larger than the lower error range. Note that $\rm FLux_{C1}$ refers to the flux (10 to 10000 keV) of the continuum spectrum except for the gaussian emission lines; $\rm FLux_{C2}$ is the flux of the continuum spectrum calculated in the gaussian 3$\rm \sigma$ energy range.}
\label{fig:Gauss_Evalution_flux}
\end{figure}

\subsection{Line and Continuum Components}

The time evolution of the spectral line and continuum components are shown in \cref{fig:Gauss_Evalution_flux}. 
We find that the ratio of line width ($\sigma$) to line central energy ($E_{\rm line}$) seems to be a constant. Fitting the data with a constant results in $10\pm 1$\% ($\chi^2$/dof is 4/7).
%From this figure 
Although the line flux ($\rm FLux_{G}$) and the continuum component flux ($\rm FLux_{C1}$ and $\rm FLux_{C2}$) may seem to have a similar evolutionary trend, the continuum component flux in the G1 time interval is not significant higher than the flux in G2 time interval, which apparently differs from the line flux evolution. The fact that the line flux has similar time evolution with the afterglow rather than the underlying continuum component flux (i.e. prompt emission) may indicate that the emission line may have some association with the general and smooth process of the jet, rather than some short-time scale irregular burst activities.

\begin{figure}[H]
\centering
\includegraphics[width=0.9\columnwidth]{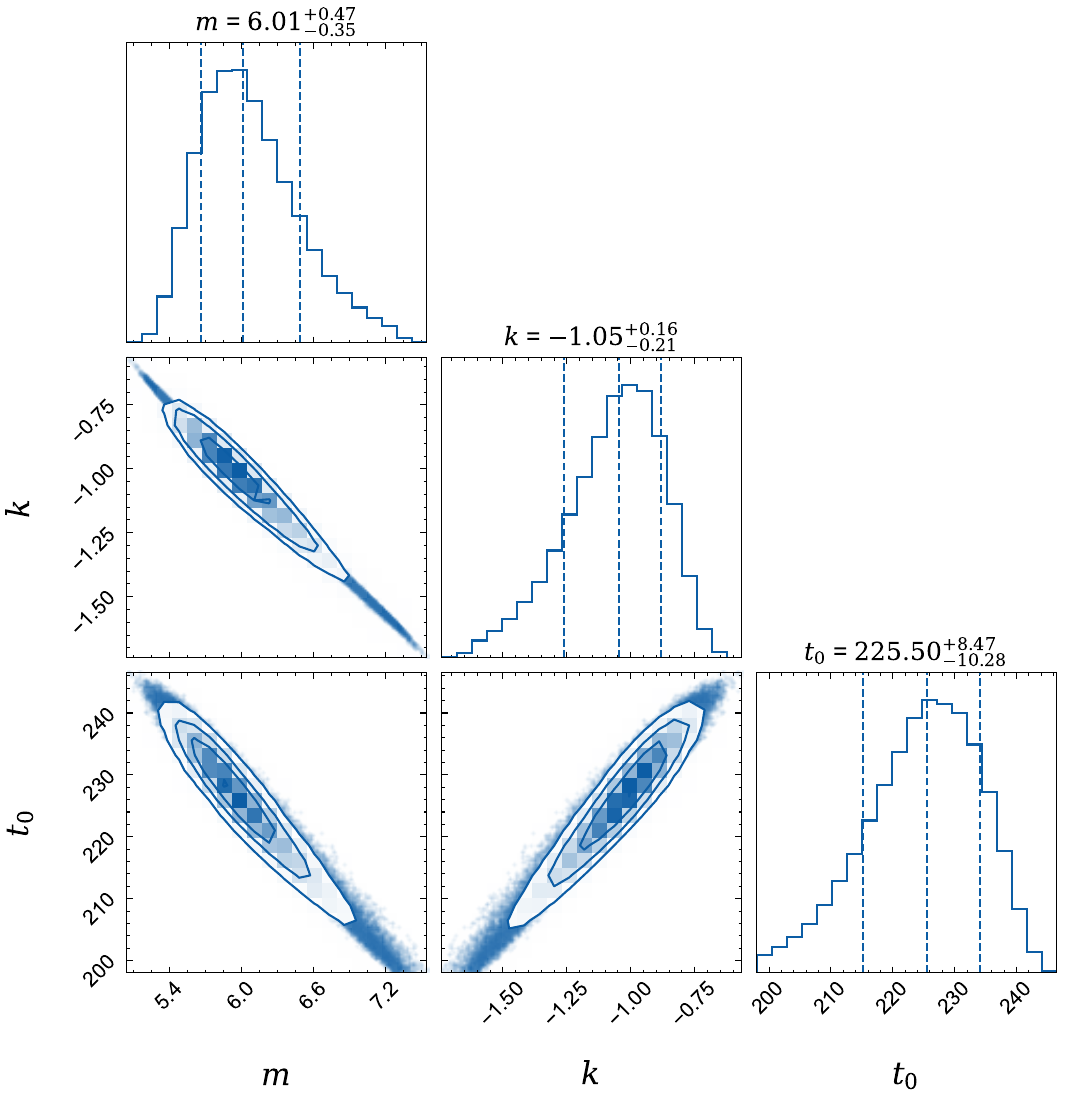}
\caption{MCMC corner of the power law fit of the time evolution of emission line central energy.}
\label{fig:corner}
\end{figure}

\begin{figure}[H]
\centering
\includegraphics[width=0.9\columnwidth]{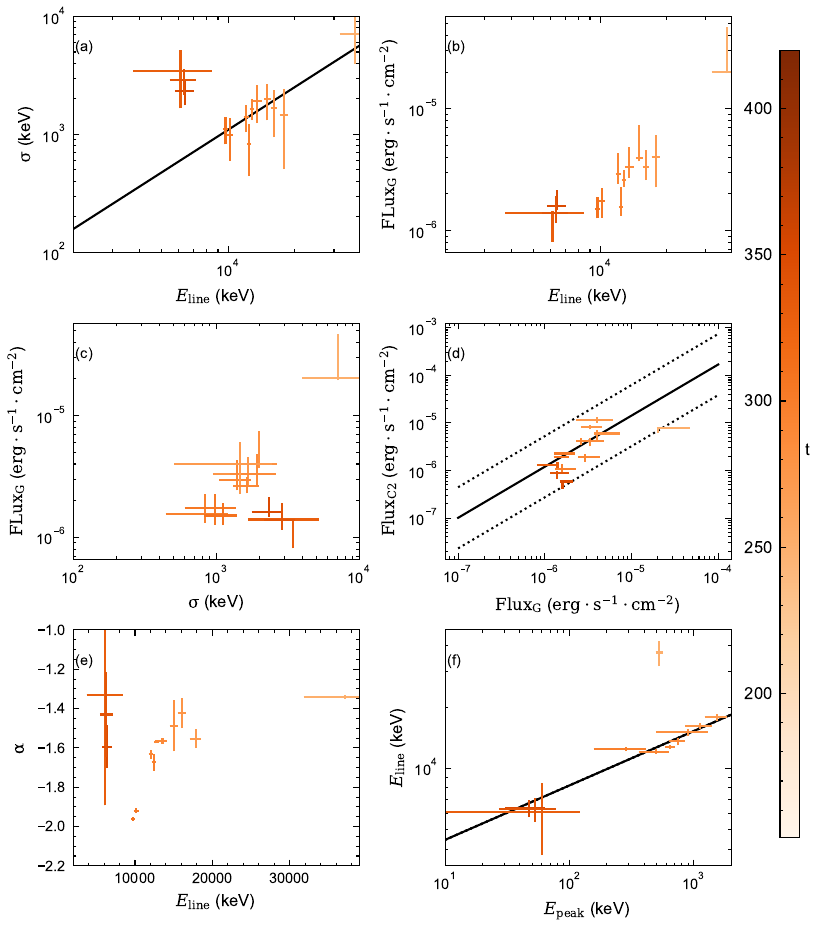}
\caption{Relationship between parameters of emission line and continuum spectral components. Different colors represent different time periods.
%, the color (from light to dark) represent the time (from small to large).
}
\label{fig:relationship}
\end{figure}

We also tried to explore the relationship between different parameters of line and continuum spectral components, and the results are displayed in \cref{fig:relationship}. 
%In this figure, the color . 
We investigate the possible relationship between each of the two parameters, including $\sigma$ vs $E_{\rm line}$, $\rm FLux_{G}$ vs $E_{\rm line}$,  $\rm FLux_{G}$ vs $\sigma$, $\rm FLux_{C2}$ vs $\rm FLux_{G}$,  $\rm \alpha$ vs $E_{\rm line}$ and $E_{\rm line}$ vs $E_{\rm peak}$.  
Several notable relationships can be found. 
In panel (a), there is an approximately linear relationship between the spectral line centre energy and $\sigma$ after removing the three outliers where the spectral lines appear very weak, and this relationship is also mentioned in \cref{fig:Gauss_Evalution_flux}. In panel (b), the spectral flux of the gaussian emission lines increase with the spectral line energy, which is easily understandable. In panel (c), $\rm FLux_{G}$ and $\rm FLux_{C2}$ show an approximately synchronous growth pattern, which means that the weaker the continuum spectrum, the weaker the gaussian line. Since the emission lines of all time intervals follow the same evolution trend while the continuum component (i.e. prompt emission) does not, the emission line and continuum components may have different origin and mechanism. However, except for the highest energy emission line (in G1 time interval), there is a linear relationship between $E_{\rm line}$ and $E_{\rm peak}$ in logarithm space, where $E_{\rm peak} = (2+\alpha)* E_{\rm c}$ is the characteristic energy (peak energy in the $\nu F_{\nu}$) of the continuum component. 
Fitting the data results in $\log_{10}(E_{\rm line})$ = $(0.27 \pm 0.06) \cdot \log_{10}(E_{\rm peak}) + (3.38 \pm 0.02)$. Implications of all these relations require further studies.

% \begin{figure*}[t]
% \centering
% \includegraphics[width=1.5\columnwidth]{Evalution_Eline_v3.pdf}
% \caption{Panel (a): evolutionary trends of lines energy over time. The blue crosses are the spectral line energies at different time periods with $1\sigma$ error bars, the green crosses are the spectral line energies after integration for the corresponding time periods with $1\sigma$ error bars, the black lines are the power-law decay fitted model line, the orange area shows the corresponding $1 \sigma$ confidence interval, and the red square dot is the energy of the absorption line with $1\sigma$ error bar. Panel (b) shows the light curves of GECAM-C (the black dots) and LHAASO (the magenta dots), and displayed with the corresponding color axes. $t_{\rm AG}$ is the afterglow onset time determined by TeV afterglow analysis \cite{lhaaso2023tera}.} 
% \label{fig:Gauss_Evalution}
% \end{figure*}

\section{Discussion and Conclusion}\label{sec:Discussion ans Conclusions}

\subsection{Time Range of Emission Lines}

Thanks to the accurate measurement of the energy spectrum by GECAM-C and higher energy band coverage by \fermi BGO detectors, we were able to search for spectral line features throughout the GRB 221009A. 
By examining both the light curves and time-resolved spectrum, we found the presence of emission lines in the time intervals \T+(246, 256) s and \T+(270, 360) s. 
% We note that the emission lines in \T+(280, 360) s were also reported by an independent work based on \fermi data alone \cite{edvige2023bright}, confirming our results in the corresponding time range. 
%
We note that our emission line energies in later time, i.e. \T+(280, 360) s, are generally consistent with the results reported in an independent work based on \fermi data only \cite{edvige2023bright}.
With the joint data of GECAM-C and \fermi, we firstly find that the emission line actually show up in much earlier time with much higher energy, i.e. 37 MeV at \T+(246, 256) s. We also revealed a remarkable power law decay behavior for both the emission line energy and line flux (see \cref{fig:Gauss_Evalution_flux}).

We note that all time intervals with detected spectral lines locate outside the two main bright peaks (220 s to 246 s and 246 s to 270 s) of GRB 221009A. The earliest time window (i.e. G1) with highest detected emission line just sits between the two main peaks where the prompt emission decreases to a relatively low level. It is also interesting to note that this time window is just after the peak time (\T+244 s) of the afterglow as measured in TeV band. %such as between the two peaks of the ME and after the end of the ME. 
However, we stress that the non-detection of spectral line during the two main bright peaks (220 s to 246 s and 246 s to 270 s) does not mean that there are no spectral lines emitted, instead, we think that the emission line could be there according to the time-evolution of emission line, but these emission lines are over-shined by the prompt emission (i.e. continuum spectral component) and thus become non-detectable, which is supported by our simulation (see \cref{sec:Spectrum Simulation} for details). 
On the other hand, the non-detection of emission line in later time (after \T+360 s) is likely caused by the low flux of the line itself, as the line flux decreases with time with a power law index of about -2 (see \cref{fig:Gauss_Evalution_flux}).

According to the power law decay of the emission line, the central energy of emission line may be much higher in earlier time, e.g., before \T+246 s and after the initial time (about \T+226 s) of this power law decay. Such a high energy emission line will be above the energy range of \fermi BGO detector, however, some other high energy instruments may be able to detect it.

\subsection{Origin of the Emission Lines}
% As shown above, we find that a series of emission lines appear in different epochs of GRB 221009A, and
% the line profiles keep the similar shape of $\sigma/E_{\rm{line}}\sim 10\%$ (where $\sigma$ is the line width and $E_{\rm{line}}$ is central energy of the line) in all time intervals, and both the line energy and line flux follow a well-defined power law decay.  These facts indicate that the emission lines should share the same origin. 

As shown above, we find that a series detections of emission line appear in different epochs of GRB 221009A, and the line profiles keep the similar shape of $\sigma/E_{\rm{line}}\sim 10\%$ (where $\sigma$ is the line width and $E_{\rm{line}}$ is central energy of the line) in all time intervals, and both the line energy and line flux follow a well-defined power law decay.  These facts indicate that the emission lines in different time intervals should share the same origin, for which we discuss some possible scenarios in the following.

%Here we suggest that the electron-positron pair annihilation is the most probable mechanism for the emission line production, as it has been already mentioned in some literatures \cite{Ioka2007ApJ,Beloborodov2010MNRAS}. Other mechanisms, such as the heavy nuclear decay, can be considered somewhere else in the future. 

Firstly, we suggest that the electron-positron pair annihilation 511 keV line is the most probable and natural production mechanism for the observed emission line \cite{Ioka2007ApJ,Beloborodov2010MNRAS}. Other mechanisms (such as the heavy nuclear decay, neutron capture line) may be also possible but we leave it for future studies. In this framework, the observed evolution of the emission line is mainly caused by the (apparent) Lorentz factor or Doppler effect of the source region of the electron-positron pairs.

The most prominent feature is that the time evolution of emission line central energy follows $E_{\rm line} \propto t^{-1}$. In order to interpret this phenomenon, one may firstly consider the high-latitude emission effect where the jet is uniform and the observed emission in later time comes from higher latitude region of the shell when the emission of the whole shell suddenly ceased or significantly reduced. This model predicts a time evolution of the observed Doppler factor $D$ with a power law index of about -1, and the observed central energy of the emission lines is proportional to $D$ (i.e. $E_{\rm line} = D m_e c^2$), thus it can nicely explain the observed $E_{\rm line} \propto t^{-1}$. However, this model also predicts that the line flux should be proportional to $D^3$ \cite{2015MNRAS.450.3549S}, thus $t^{-3}$.
%Our observations indicate that $E_{\rm line} \propto t^{-1}$, implying that the flux should also be proportional to $t^{-3}$. 
But our observations show that the measured line flux is proportional to $t^{-2}$, which is inconsistent with the expectation of the high latitude effect.
To alleviate this contradiction, one may invoke other factors in the high latitude emission scenario, such as the down-Comptonization effect \cite{2020ApJ...900...10L} and the non-uniform structure of the jet. 

However, the similar time evolution between the emission line and the TeV afterglow inspired us to consider a more probable scenario where the evolution of the emission line is caused by the global evolution of the jet itself, just like the afterglow.
We consider that the source of the emission line is some dense plasma clumps with electron-positron pair creation and annihilation, and this source region travels along with the global jet propagation. In such scenario, the line central energy directly reflects the (apparent) Lorentz factor of the jet. According to the evolution of the emission line energy, we drive that evolution of
the bulk Lorentz factor of the jet is $\Gamma\propto t^{-1}$. This means that this scenario allows us to directly measure, for the first time, the speed of the GRB jet and characterize the evolution of the bulk Lorentz factor of the jet when the jet is just accelerated to and still in the highly relativistic regime. If this is true, we find that our result ($\Gamma\propto t^{-1}$) shows that the jet dynamic evolution of GRB 221009A is much faster than the general cases predicted by theoretical modeling \cite{1998ApJ...497L..17S}. 

% Our observation of the emission lines allow us, for the first time, to directly measure the speed of the GRB jet and characterize the evolution of the bulk Lorentz factor of the jet
% %in the prompt emission 
% when the jet is just accelerated to and still in the highly relativistic regime. According to the evolution of the emission line energy, we drive that evolution of
% the bulk Lorentz factor of the jet is $\Gamma\propto t^{-1}$. This result shows that the jet dynamic evolution of GRB 221009A is much faster than the general cases derived from theoretical modeling \cite{1998ApJ...497L..17S}. \textbf{If we consider the effect from the high-latitude emission, we obtain $L_{\rm{line}}/E_{\rm{line}}\propto t^{-2}$. Then, we obtain $\Gamma\propto t^{-1}$ when we suggest $L_{\rm{line}}\propto \Gamma^3$ and $E_{\rm{line}}\propto \Gamma$. The time evolution of the emission line might be affected by the down-Comptonization \cite{2020ApJ...900...10L}}.

In this scenario, we can extrapolate the jet bulk Lorentz factor to \T+230 s when the TeV afterglow just emerged, and estimate the Lorentz factor $\Gamma={485}_{-276}^{+245}$, 
which is fairly consistent with the initial bulk Lorentz factor ($\Gamma\approx440$) estimated from the TeV afterglow observation \cite{lhaaso2023tera}. 
%(SNZ comment: the above sentence is confusing. I suggest to break this sentence into several sentences, by explaining clearly: 1) when was the first TeV emission detected by LHAASO? (228 or 382 or 230 s? With respect to what time?) 2) What is the value of extrapolated $\Gamma$ to the time when the first TeV emission was detected by LHAASO? 3) What is the value of the initial bulk Lorentz factor estimated by the onset of the afterglow? 4) Comparison between the above two values of $\Gamma$.)} 
This consistency not only confirms that it is reasonable to infer the speed of the jet using the emission line central energy but also provides an additional support to this scenario. 
%\Tred{Moreover, our results shows that a highly relativistic jet indeed presented in this GRB and the bulk Lorentz factor may be even higher if it strictly follows the power law decay in earlier time, however, there should be an upper limit in the bulk Lorentz factor otherwise it will go to infinite at the initial time (around \T+226 s) (SNZ comment: I suggest to remove this.)}.

In principle, the initial time ($t_0$=${226}^{+8}_{-10}$~s ) of the power law decay of emission line corresponds to the time when the jet starts to decelerate. We note that this initial time is not only coincident with the rising phase of the first main peak of the prompt emission observed by GECAM-C (see \cref{fig:Gauss_Evalution}), but also remarkably consistent with the the afterglow onset time ($t_{\rm AG}$ = ${225.7}^{+2.2}_{-3.2}$~s ) obtained by fitting the TeV afterglow light curve \cite{lhaaso2023tera}. 
%Furthermore, by using \cref{equ:exp_model} to fit the flux of LHAASO, we are also get an initial time of ${226}^{+2}_{-2}$~s (refer to \cref{sec:LHAASO Flux Fit}).

Comparing the time evolution of the emission line flux (${\rm FLux_G} \propto t^{-2}$) and emission line energy ($E_{\rm line} \propto t^{-1}$), we find that the line flux evolution is well explained by a Doppler boosting ($\Gamma$ of the jet) of a constant flux of 511 keV annihilation line in the co-moving frame. Note that this Doppler effect will not change the ratio between the line width and the line central energy, which is also in a good concordance with the observation ($\sigma/E_{\rm{line}}\sim 10\%$ in all time intervals) mentioned above. Indeed, the emission line width is likely caused by the different Doppler factors of the emitting clumps and/or the temperature of the pair plasma inside the clumps.

In order to produce the emission lines with the flux of $(1.0-4.0)\times 10^{-6}~\rm{erg~cm^{-2}~s^{-1}}$ (\cref{Spectrum fitting results with Gauss Component}), we estimate that the plasma clump should have a large number density of about $(1.2-2.2)\times 10^{13}~\rm{cm^{-3}}$ \cite{svensson1982pair}. In such case, the co-moving volume is assumed to be a cone with the half opening angle of $\theta\approx0.6$ degree, and the radius of the emission region is about $1.0\times 10^{15}$ cm \cite{lhaaso2023tera,HXMT-GECAM:GRB221009A}. 

\section{Summary}\label{sec:Summary}
In this work, we carried out a comprehensive spectral analysis of the record-breaking GRB 221009A jointly with GECAM-C and \fermi data, focusing on the search for emission and absorption lines. By taking advantages of these two instruments, we were able to probe the spectrum of the full course of this GRB including the bright part regime, % of the prompt emission and flare phase, 
allowing us to accurately measure the continuum spectral component and reveal unprecedented details of emission lines. For the first time, we found that the maximum central energy of emission line is up to about 37 MeV. This measurement was not reported before, because it occurred during the most bright part of this GRB, where many gamma-ray monitors (except for GECAM-C) suffered severe instrumental effects.
 More importantly, we revealed a remarkable time evolution of the emission line in the form of a power law decay, which resembles the behavior of the TeV afterglow. %thus the GRB dynamic evolution.
%TeV afterglow evolution observed during the same time range of the emission line. 
We discussed some possible scenarios to interpret our observation results. First, we disfavored the simple high latitude emission effect for an uniform jet based on the observed evolution of central energy and flux of the emission line. %because the observation of spectrum line ${\rm FLux_G} \propto t^{-2}$. 
%This implies the involvement of more physical processes or a structured jet.
Instead, we suggest that the observed emission line could be reasonably explained by the blue-shifted electron-positron pair annihilation 511 keV line emit by some dense clumps moving together with the relativistic jet. This scenario allows us, for the first time, to directly measure the speed of the jet during the high relativistic regime, and the maximum measured central energy (about 37 MeV) of the emission line corresponds to the jet Lorentz factor of $\Gamma=84 \pm 14$. 
Interestingly, in this scenario the flux of the electron-positron pair annihilation 511 keV line in the co-moving frame seems to be constant during the time period of detection of the emission line.  
However, more theory work is required to give a solid and complete interpretation for our discoveries, which may help to probe unprecedented details on the hydrodynamics and compositions of the relativistic jet in GRB.

%physics of the jet and GRB physics.

%The physical processes, such as the component loading,
%the heavy element nucleosynthesis,
%and the magnetization,
%are further revealed by the observation.  

%Discuss the trend in the evolution of $E_{\rm line}$ over time, possibly corresponding to the evolution of the lorentz factor, etc. Possibly a direct spectroscopic measurement of the deceleration process of the jet after hitting the surrounding medium.

% \newpage
\bibliography{reference}
% \bibliographystyle{aasjournal}

%%%%%%%%%%%%%%%%%%%%%%%%%%%%%%%%%%%%%%%%%%%%%%%%%%%%%%%
%%% Acknowledgements. ??§Ý
%%%%%%%%%%%%%%%%%%%%%%%%%%%%%%%%%%%%%%%%%%%%%%%%%%%%%%%
\Acknowledgements{This work is supported by 
the National Key R\&D Program of China (Grant No. 2021YFA0718500, % HXMT & GECAM
2023YFE0101200), % JM
the Strategic Priority Research Program of Chinese Academy of Sciences (Grant No. XDA15360102, XDA15360300, XDA15052700), % GECAM
the National Natural Science Foundation of China 
(Grant No. 12273042, % xiong shaolin
61234003, 61434004, 61504141, 11673062, 12393813, % JM
2333007, 12027803 %SNZ
and 12303045) %caice
and CAS Interdisciplinary Project (Grant No. KJZD-EW-L11-04). % JM
The GECAM (Huairou-1) mission is supported by the Strategic Priority Research Program on Space Science (Grant No. XDA15360000) of the Chinese Academy of Sciences.
Ji-Rong Mao is supported by the Yunnan Revitalization Talent Support Program (YunLing Scholar Award).
We appreciate the development and operation teams of SATech-01 and GECAM.
We acknowledge the public data and software from {\it Fermi}/GBM. 
We appreciate helpful discussions with Xiang-Yu Wang, Binbin Zhang, Tao An, Bing Zhang, He Gao, Hua Feng, Yong-Feng Huang, Zi-Gao Dai, Yun-Wei Yu and Ti-Pei Li.
Author contributions: Shao-Lin Xiong led this study and GECAM-C observation and data analysis. Yan-Qiu Zhang is the major contributor of data analysis. Ji-Rong Mao, Shuang-Nan Zhang, Shao-Lin Xiong, Xi-Lu Wang, Ming-Yu Ge, Shu-Xu Yi contributed the theory interpretation. Shao-Lin Xiong, Yan-Qiu Zhang, Ji-Rong Mao, Shuang-Nan Zhang, Xi-Lu Wang contributed the writing of manuscript. All authors contributed to the GECAM-C mission and this work.} 

%%%%%%%%%%%%%%%%%%%%%%%%%%%%%%%%%%%%%%%%%%%%%%%%%%%%%%%
%%% Conflict of interest. ????????????
%%%%%%%%%%%%%%%%%%%%%%%%%%%%%%%%%%%%%%%%%%%%%%%%%%%%%%%
\InterestConflict{The authors declare that they have no conflict of interest.}

%%%%%%%%%%%%%%%%%%%%%%%%%%%%%%%%%%%%%%%%%%%%%%%%%%%%%%%
%%% Supplements. ????????, ????
%%%%%%%%%%%%%%%%%%%%%%%%%%%%%%%%%%%%%%%%%%%%%%%%%%%%%%%
%\Supplements{}

%%%%%%%%%%%%%%%%%%%%%%%%%%%%%%%%%%%%%%%%%%%%%%%%%%%%%%%
%%% Reference section. ?¦Ï?????
%%% citation in the content using "some words~\cite{1,2}".
%%% ~ is needed to make the reference number is on the same line with the word before it.
%%%%%%%%%%%%%%%%%%%%%%%%%%%%%%%%%%%%%%%%%%%%%%%%%%%%%%%

\normalem

%%%%%%%%%%%%%%%%%%%%%%%%%%%%%%%%%%%%%%%%%%%%%%%%%%%%%%%
%%% Appendix sections. ??????, ????
%%%%%%%%%%%%%%%%%%%%%%%%%%%%%%%%%%%%%%%%%%%%%%%%%%%%%%%
\begin{appendix}
%\section{Name}

%\end{appendix}

%\begin{appendices}
%\section{Appendix}
%\end{appendices}
%\appendix

%\appendix

% \newpage
\renewcommand{\thesection}{Appendix}

\section{}

\subsection{\label{sec:Fermi check} Data Check for Fermi/GBM}

\subsubsection{\label{sec:Saturation} Data Saturation}
\begin{figure}[H]
\centering
\includegraphics[width=\columnwidth]{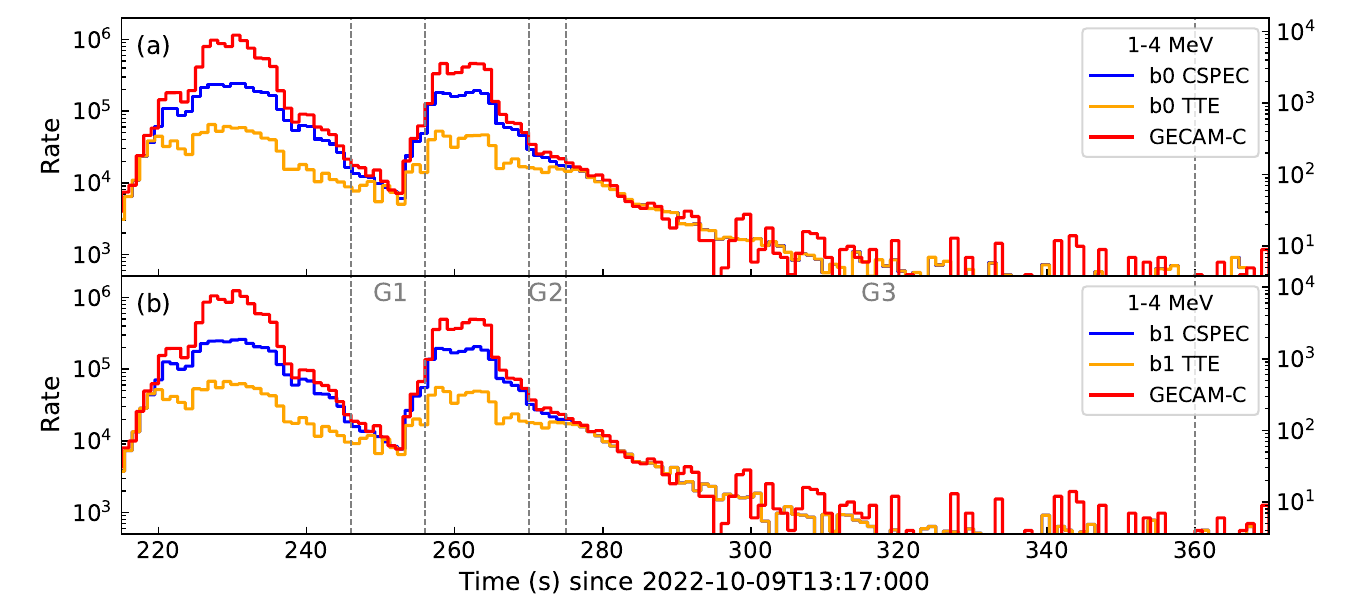}
\caption{Comparison of the lightcurves generated by \fermi TTE, Cspec data products and GECAM-C. The two panels show the lightcurves of b0 and b1 detectors respectively; note that we use a factor to correct for the difference in effective area between \fermi and GECAM-C. The results show that Cspec suffers relatively little data loss relative to the GECAM-C data over the two time periods we analyzed.}
\label{fig:GBM_data_Saturation}
\end{figure}

These two time intervals G1 \T+(246, 256) s and G2 \T+(270, 275) s locate in \fermi BTI time period, during which \fermi suffers from data saturation. As shown in \cref{fig:GBM_data_Saturation}, we can see that Cspec data lost very few data compared to GECAM-C, which is consistent with the constant factor of $1.05 \pm 0.08$ of the joint fit of GBM and GECAM-C data.

\subsubsection{\label{sec:Background} Background}
\begin{figure}[H]
\centering
\includegraphics[width=\columnwidth]{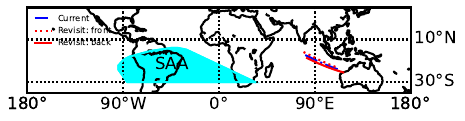}
\caption{\Fermi's geographic locations in the time period \T+(0,600) s and the revisit orbits. The blue dotted-dashed line indicates the geographic coordinates of the satellite in its current orbit, the red dotted line indicates the revisit orbit of the front and the red line indicates the revisit orbit of the back. The difference between the two is very small, which means that the two orbits have almost the same space particle environment.}
\label{fig:GBM_lonlat}
\end{figure}

\begin{figure}[H]
\centering
\includegraphics[width=0.8\columnwidth]{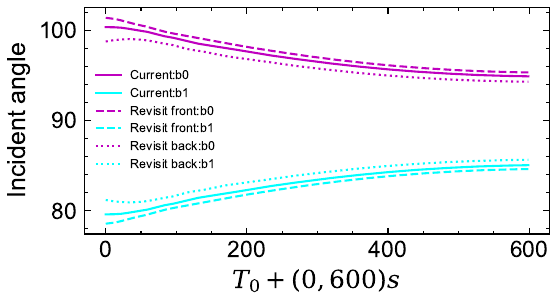}
\caption{Incident angles of b0 and b1 in the time period \T+(0,600) s and the revisit orbits. The solid line represents the current orbit, the dashed line represents the revisited orbits of the front and the red dotted lines indicate revisit orbits of the back, the change in incidence angle is very small, which means that the attitude change of \fermi is very small in the two-orbit case.}
\label{fig:GBM_Incident_angle}
\end{figure}

First, the impact of the background will be much less significant during periods when the burst is particularly intense. Secondly, as shown in \cref{fig:LC_For_ME_GL_GBM}, the light curves of various energy bands indicate that the background subtraction is quite reliable.
Thirdly, as shown in \cref{fig:GBM_lonlat,fig:GBM_Incident_angle}, the geographic position and attitude of the \fermi satellite remained almost constant. %Additionally, over the course of a single day, the overall state of the satellite undergoes only minor fluctuations. 
In summary, we can utilize the method of revisit orbits to estimate the background accurately.

\subsubsection{\label{sec:Energy} Selection of Energy Range}

\begin{figure}[H]
\centering
\includegraphics[width=0.9\columnwidth]{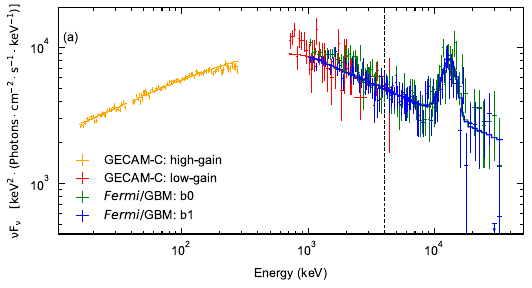}
\caption{The comparison of the spectra of \fermi b0 and b1, where systematically different behaviour can be seen. The black dashed line marks 4 MeV.}
\label{fig:Specfit-examples}
\end{figure}

In our data analysis, we do not use data from NaI detectors for two reasons. First, the high-gain of GECAM-C can cover the energy range of 15-300 keV. Second, there is inconsistency among NaI detectors mentioned before \cite{Fermi:GRB221009A, edvige2023bright}. As show in \cref{fig:Specfit-examples}, there is also a clear inconsistency between the data of the two BGO detectors below 4 MeV. So we just choose the data above 4 MeV when analyzing the BGO data. Fortunately GECAM-C low-gain data can cover up to 5.5 MeV. Note that the available energy band also changes with the counts rate. The lower the count rate, the lower energy limit we can select to use. %Therefore, we speculate that the inconsistency may be due to the impact of high count rate, which alters the spectral profile.

\vspace{-0.5cm}

\begin{figure}[H]
\centering
\includegraphics[width=0.75\columnwidth]{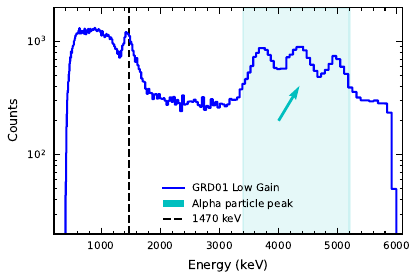}
\caption{Energy spectrum drawn from cumulative GECAM-C data. The black dashed line corresponds to the characteristic peak at 1470 keV of $^{138}\mathbf{La}$. The cyan shaded area represents the $\mathbf{\alpha}$ particle peak. The blue arrows point to the $\alpha$ particle peaks.}
\label{fig:Characteristic_peak}
\end{figure}

\subsection{\label{sec:GECAM-C check} An interesting time interval}
\begin{figure}[H]%h-t-b-p
\centering
\includegraphics[width=0.85\columnwidth]{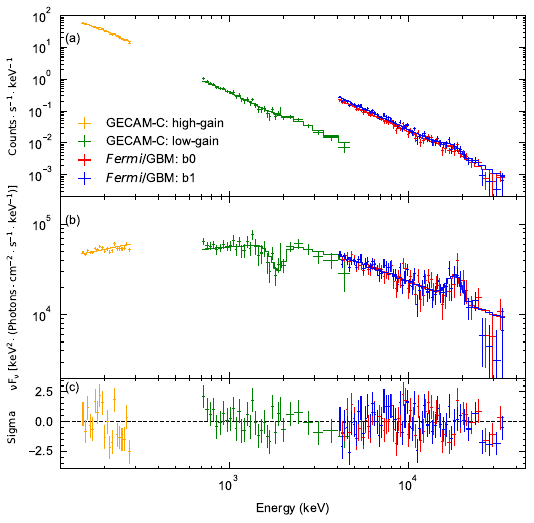}%"scale"width=1.0\columnwidth"
\caption{The spectral fitting result with \textit{band*gabs+gauss} spectrum model within the time interval \T+(270, 275) s. The definition of the colors are the same as \cref{fig:32MeV}. Panel (c) is the residuals in terms of sigmas with error bars of size one. As shown in this figure, there are two components: a suspected absorption line and a gaussian emission line. Note that for GECAM-C, the data below 150 keV are ignored due to the impact of the high count rate.} 
\label{fig:Two-components}%{}"fig:example1"\ref{fig:example1}
\end{figure}

We find there is an interesting time interval, \T+(270, 275) s, where one emission line and one marginal absorption line features are seen, as shown in \cref{fig:Two-components}. The emission line has its central energy and width of $17.84 \pm 0.76$ MeV and $1.47 \pm 0.96$ MeV respectively. The chance probability value is $3.90\times10^{-3}$, and the Gaussian-equivalent significance is 2.89 $\sigma$. Although the significance is not high, this line energy follows perfectly the same evolution as the other emission lines before and after this time interval; thus we consider the emission line in this time interval is true and the low significance is caused by the relative brightness of the line and continuum component of the spectrum. 

Intriguingly, the marginal absorption line also follows the Gaussian shape with its 
central energy and width of $1.87 \pm 0.19$ MeV and $144 \pm 77$ keV. We note that this is only absorption feature we found throughout the burst and it has a similar line with to line central energy ratio as the emission lines. However, we caution that the significance of this absorption line is very low (about 1.5 $\sigma$ without considering the trial numbers during the search). In the following, we still make some discussions on it in view of the importance of spectral line in GRB spectrum.

Because this marginal absorption line was detected in GECAM-C data, we carefully checked the background spectrum of GECAM-C to investigate any possible systematic effect of background subtraction. As can be seen in \cref{fig:Characteristic_peak}, 
we accumulated one-hour \textbf{EVT} data of GECAM-C and plotted the energy spectrum to display the characteristic peaks of GECAM-C with low gain data. The spectrum is very smooth without significant line features from 1.8 MeV to 3 MeV. The flat structure means that we cannot subtract a bump or a peak from the full spectrum, which can help us to rule out the background subtraction origin of the detected absorption line at 1.87 MeV.
Furthermore, since the background only constitutes about 10\% of the full spectrum during the time period \T+(270,275) s, which means that any variation in the background spectrum will not sufficiently produce a gaussian absorption line structure.
There are some line features outside of the above energy range, e.g. the $^{138}\rm La$ intrinsic peak at 1470 keV, the alpha particle peaks from 3.5 MeV to 5 MeV. 
We emphasize that these lines are produced by well-understood mechanisms related to the GECAM-C detector and will not change their energy \cite{zhangperformance}. These analysis suggest that this absorption line should not come from instrumental effects, if this line is real.

% Although the significance of this absorption line is low (about 1.5 $\sigma$), we note that this is the only absorption line feature found in all time intervals during the prompt emission stage of GRB 221009A and this absorption feature just appears when the emission line emerge after the second bright peak in the main burst (see \cref{fig:Gauss_Evalution}). 

 % There are some line features outside of the above energy range, e.g. the $^{138}\rm La$ intrinsic peak at 1470 keV, the alpha particle peaks from 3.5 MeV to 5 MeV. 
%This contributes to the existence of the absorption line.

\subsection{\label{sec:Sigcal} Significance Calculation of Spectral Line}

We aim to demonstrate the significance of gaussian spectral lines through simulations. This is accomplished as follows:

% First, we simulate the data by generating the total and background spectra. To conveniently achieve this, we utilize xspec's fkeit to create a continuous spectrum. Next, we fit the model using both the continuous spectrum alone and the continuous spectrum combined with the Gaussian component. Finally, we calculate the significance using the likelihood ratio, which is represented by the difference in Δcstat.
\begin{itemize}
    \item [a)] 
    First, we use \texttt{xspec}'s \textbf{fakeit} to simulate the total and background spectral data.
    \item [b)] 
    Next, we fit the simulated data using the continuum spectrum and the continuum spectrum plus gaussian components, respectively.
    \item [c)] 
    Finally, we calculate the significance using the likelihood ratio, which is represented by the difference in $\rm \Delta cstat$.
\end{itemize}
\begin{equation}  \begin{aligned}  
LR & = 2 \cdot \ln \frac{L_{\rm com+gauss}}{L_{\rm com}} \\
   & = -2 \ln L_{\rm com} - (-2 \ln L_{\rm com+gauss}) \\
   & = cstat_{\rm com} -cstat_{\rm com+gauss} \\
   & = \Delta cstat \\
   & \sim \chi^2(x,dof=3)
\label{eq:LR test}
\end{aligned}
\end{equation}
where $L_{\rm com+gauss}$ denotes the likelihood value using the continuum spectrum plus gaussian fitting, $L_{\rm com}$ denotes the likelihood value of fitting using only the continuum spectrum, and the three degrees of freedom of chi-squared are derived from the three free parameters of gaussian or absorption.

 We use the likelihood ratio test to give significance: the original hypothesis $\mathcal{H}_0$ is that the data can be well described by a continuum spectrum, the alternative hypothesis $\mathcal{H}_1$ is that an additional spectral component needs to be added on top of the continuum spectrum, and the expression of the equation is shown in the 
\cref{eq:LR test}. We performed several simulations to analyze the distribution of $ N_{\rm sim}(>\Delta {\rm cstat}) \mid N_{\rm sim}$ with $\rm \Delta cstat$, as shown in \cref{fig:Significance}. The results show that the simulated distribution is consistent with the theoretical chi-squared distribution. Finally, we can assess the significance of the additional component by analyzing the $\rm \Delta cstat$ values recorded in \cref{Spectrum fitting results with Gauss Component,Spectrum fitting results without Gauss Component} during the reality fitting process. The $p$-value is calculated based on the actual simulation results. If a sufficient number of simulations cannot be achieved, we report \textless 1/N, and the results inferred from the chi-square distribution are also calculated. We note that the likelihood ratio does not always obey the chi-square distribution when comparing models that are complex (i.e. with emission line).

% Eventually we found the highest significant time period except \T+(246, 256) s is \T+(280, 300) s with a value of 7.83$\sigma$.

One thing to note, when calibrating the likelihood ratio test distribution by simulation, the extra parameters introduced in alternative hypothesis, i.e. the intensity, location and width of the line, are not fixed to predetermined values but are allowed to vary freely during the fit. This is a standard setup when performing the simulation \cite{2002ApJ...571..545P}. Therefore there is no trial number issue for these extra parameters, for example, the location of the line.

\begin{figure}[H]%h-t-b-p
\centering
\includegraphics[width=0.95\columnwidth]{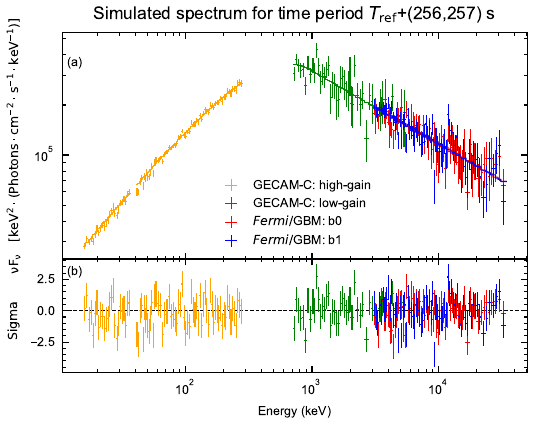}%"scale"width=1.0\columnwidth"
\caption{Simulation of the energy spectrum in the time period \T+(256, 257) s. The data points in panel (a) are simulated using the \textbf{fakeit} tool with the \textit{band+gauss} model.The solid line in panel(a), also known as the model line, is the result of only the continuous spectrum. Panel (b) is the residuals in terms of sigmas with error bars of size one. } 
\label{fig:Simulated_spectrum}%{}"fig:example1"\ref{fig:example1}
\end{figure}

\subsection{\label{sec:Spectrum Simulation} Spectrum Simulation}
We selected the time period  \T+(256, 257) s for the simulation, for which no spectral line is detected. This time period is not as bright as the two main peaks, but still considerably brighter than that of \T+(246, 256)s. 

The simulation proceeds as follows:
\begin{itemize}
    \item [a)] 
    First, we use \texttt{xspec}'s \textbf{fakeit} to simulate the total and background spectral data using the \textit{band+gauss} model. The parameters of the continuous spectrum \textit{band} are derived from a separate fit to GECAM-C. The gauss parameters are extrapolated from the power-law decay trend of $E_{\rm line}$ and the ratio of 10 \%. Therefore, at the midpoint of time \T+265.5 s, the energy and width of the spectral line are 28 MeV and 2.8 MeV respectively. In addition, for the norm we chose 0.2 (which is already large relative to 0.13 for the time period \T+(280,300) s, where the significance is very high).
    \item [b)] 
    Next, we choose the model to be \textit{band} while fixing the parameters of the continuous spectrum to be the model values used for simulation.
    \item [c)] 
    Finally, plot the eefv and the residuals as shown in \cref{fig:Simulated_spectrum}.
\end{itemize}

As shown in \cref{fig:Simulated_spectrum}, the spectral line can not be identified clearly for this time interval where the continuum component is not very bright yet. Thus, the non-detection of the spectral line during the two main peaks are well expected.

\subsection{\label{sec:LHAASO Flux Fit} Fit of the TeV light curve}
\begin{figure}[H]%h-t-b-p
\centering
\includegraphics[width=0.95\columnwidth]{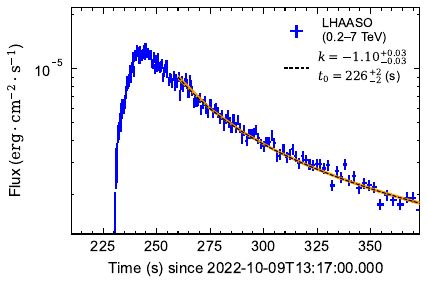}%"scale"width=1.0\columnwidth"
\caption{LHAASO's flux data fitted using \cref{equ:exp_model}. the black lines are the power-law decay fitted model line, the orange area shows the corresponding $1 \sigma$ confidence interval.} 
\label{fig:LHASSO_fit}%{}"fig:example1"\ref{fig:example1}
\end{figure}

\begin{figure}[H]%h-t-b-p
\centering
\includegraphics[width=0.95\columnwidth]{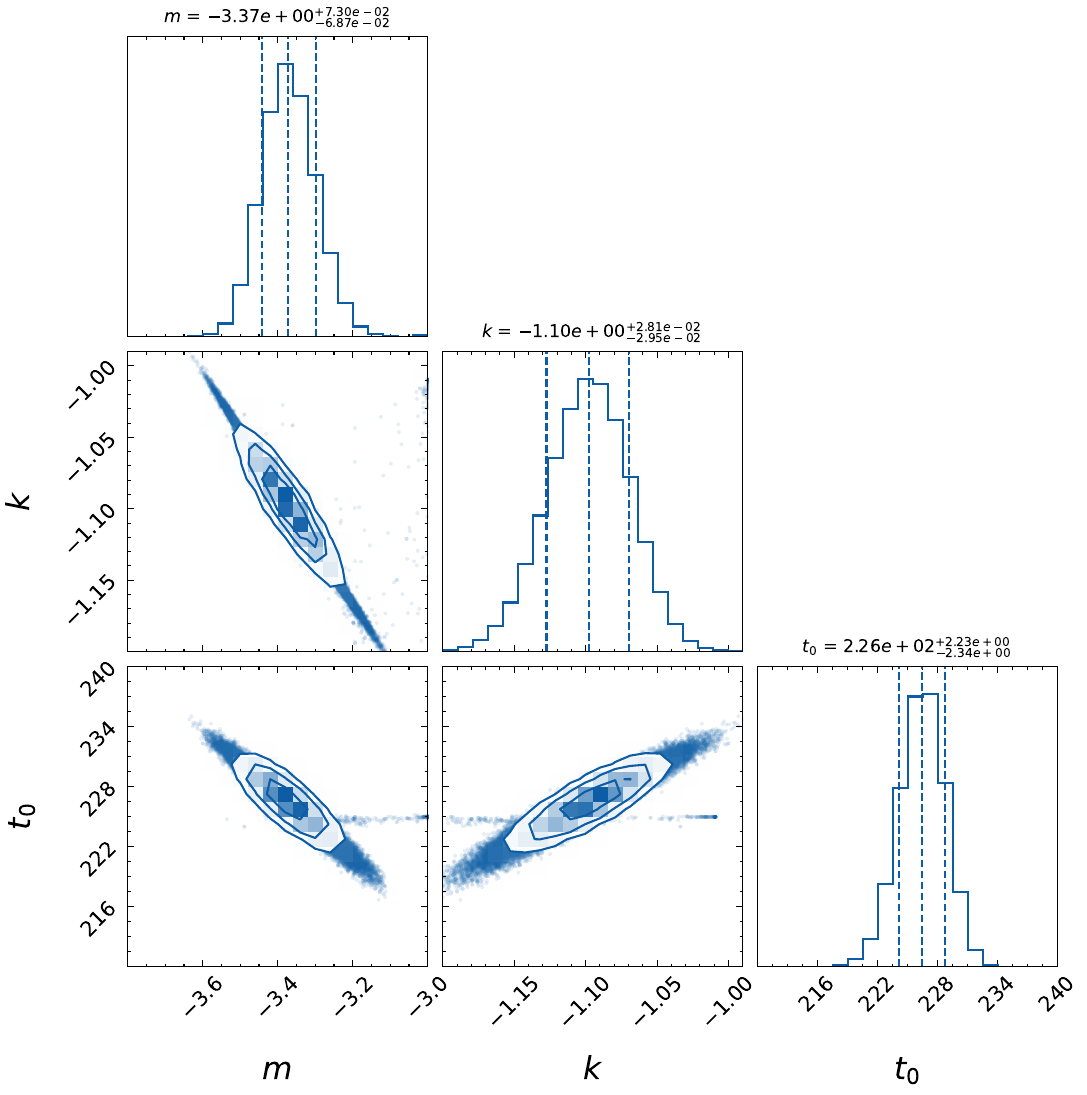}%"scale"width=1.0\columnwidth"
\caption{The MC corner result of the power law decay fit with LHAASO flux datas. All parameters are well constrained.} 
\label{fig:LHASSO_fit_corner}%{}"fig:example1"\ref{fig:example1}
\end{figure}

Just like fitting the $E_{\rm line}$ evolution over time, we also fit the Tev flux data from LHAASO using the same formula (equ. 6) as we fit $E_{\rm line}$. The results are displayed in \cref{fig:LHASSO_fit}, with the fitted MC corner plots shown in \cref{fig:LHASSO_fit_corner}. The results are $ m = {-3.37}^{+0.07}_{-0.07}$, $ k = {-1.10}^{+0.03}_{-0.03}$ and $ t_0 = {226}^{+2}_{-2}$~s. It is worth noting that, to avoid the influence of other segments in TeV afterglow, we chose a fitting time period of \T+(260,700) s. 
%Based on the results, we can find that the obtained results and the parameters of the $E_{\rm line}$'s evolution over time are consistent within the error margins.

\subsection{\label{sec:Spectrum Fitting Results} Spectrum Fitting Results}
\begin{figure*}
% \centering
% \hspace{15mm}
 \begin{minipage}{0.32\linewidth}
 	\vspace{3pt}
   	\centerline{\includegraphics[width=\textwidth]{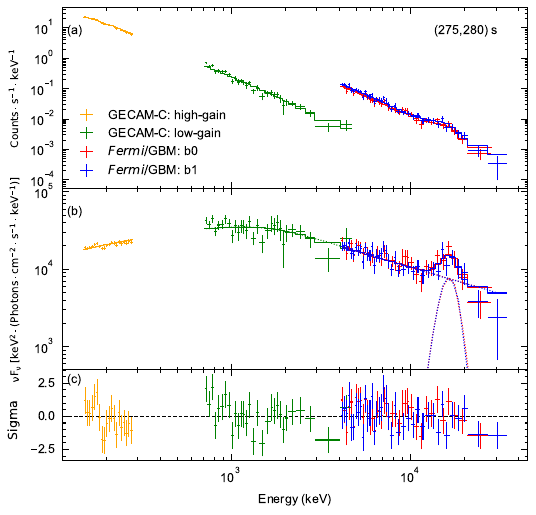}}     
        \vspace{3pt}
 	\centerline{\includegraphics[width=\textwidth]{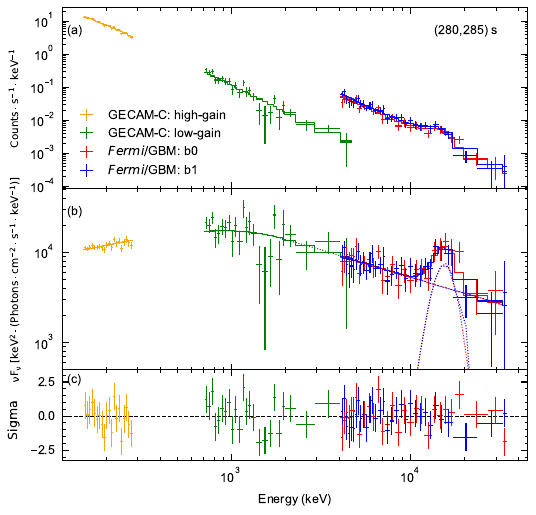}}
 	\vspace{3pt}
 	\centerline{\includegraphics[width=\textwidth]{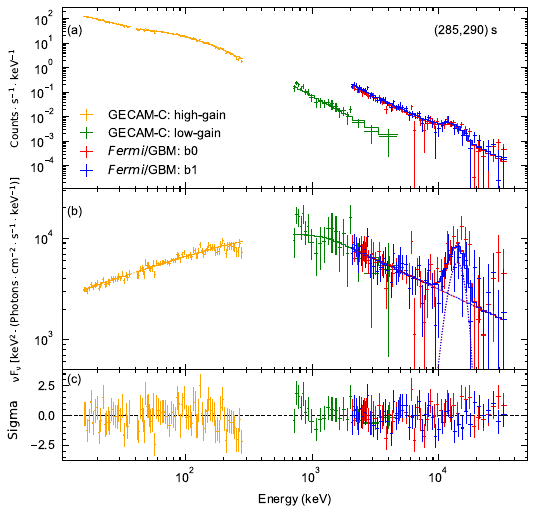}}
 	\vspace{3pt}
 	\centerline{\includegraphics[width=\textwidth]{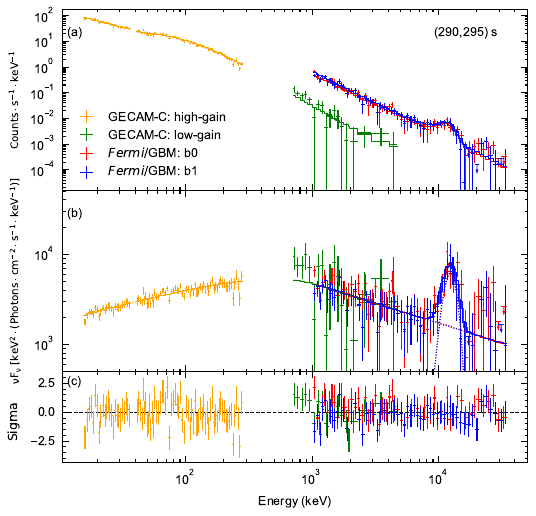}}
  	\vspace{3pt}
 \end{minipage}
 \begin{minipage}{0.32\linewidth}
 	\vspace{3pt}
        \centerline{\includegraphics[width=\textwidth]{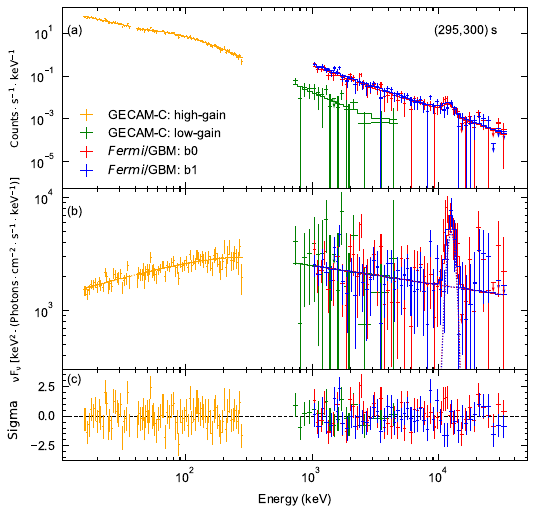}}
 	\vspace{3pt}
	\centerline{\includegraphics[width=\textwidth]{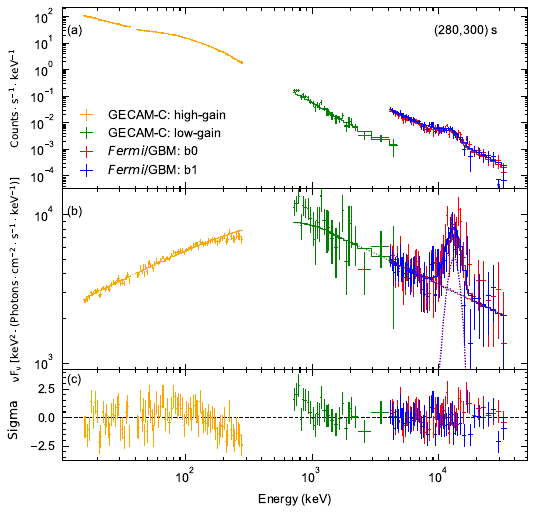}}
	\vspace{3pt}
 	\centerline{\includegraphics[width=\textwidth]{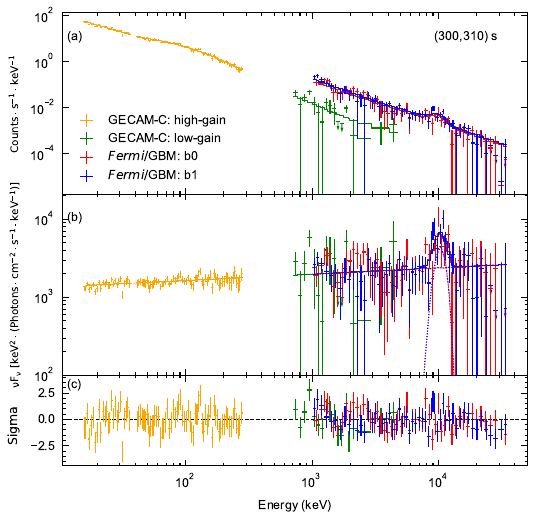}}
	\vspace{3pt}
	\centerline{\includegraphics[width=\textwidth]{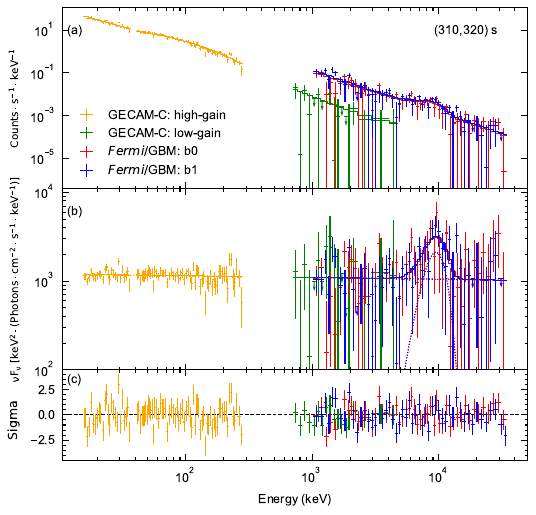}}
	\vspace{3pt}
 \end{minipage}
 \begin{minipage}{0.32\linewidth}
	\vspace{3pt}
 	\centerline{\includegraphics[width=\textwidth]{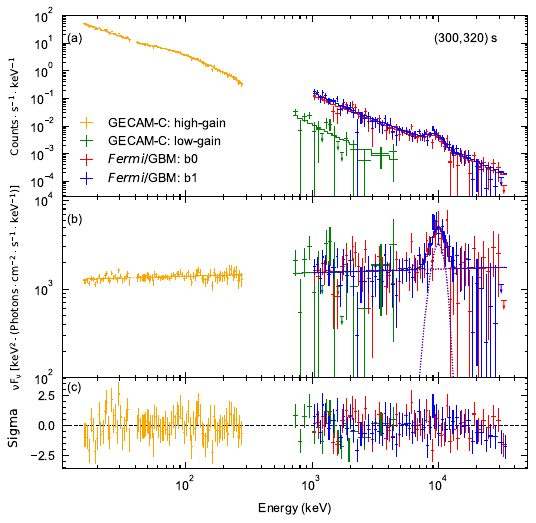}}
	\vspace{3pt}
	\centerline{\includegraphics[width=\textwidth]{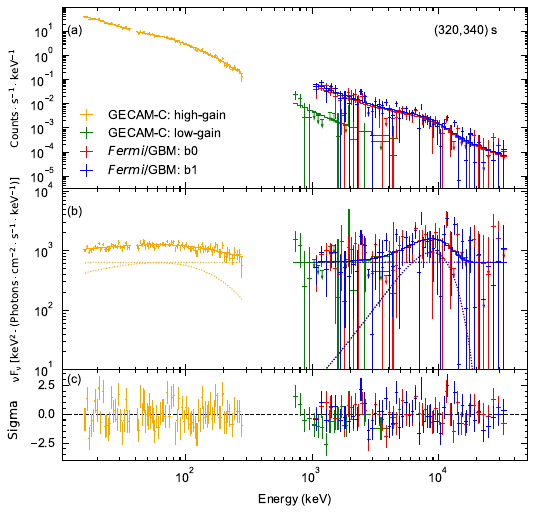}}
	\vspace{3pt}
	\centerline{\includegraphics[width=\textwidth]{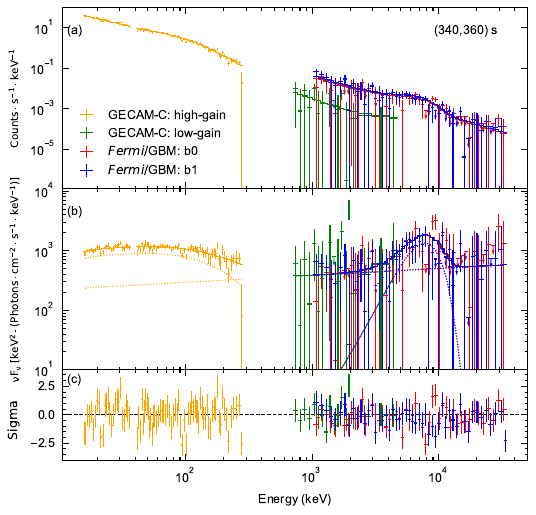}}
	\vspace{3pt}
	\centerline{\includegraphics[width=\textwidth]{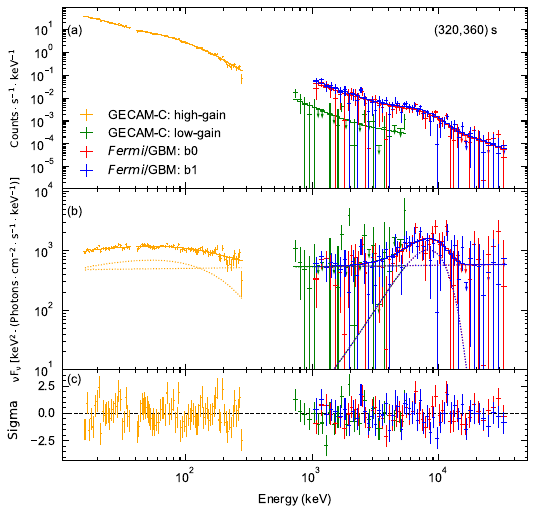}}
	\vspace{3pt}
\end{minipage}

\caption{The spectrum fitting results for each time interval, with each subplot divided into three panels: panel (a) displays the count spectrum, panel (b) shows the $\nu F_{\nu}$ distribution and panel (c) shows the residual distribution.}

\label{fig:fitting}
\end{figure*}

\begin{figure*}
	
 \begin{minipage}{0.32\linewidth}
 	\vspace{3pt}
   	\centerline{\includegraphics[width=\textwidth]{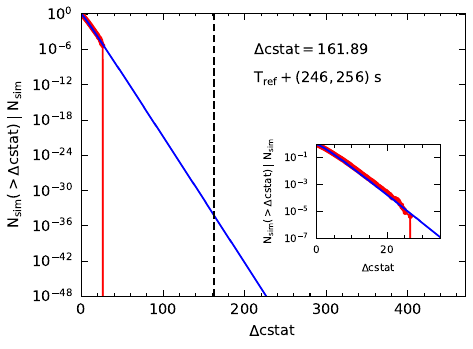}}
        \vspace{3pt}
 	\centerline{\includegraphics[width=\textwidth]{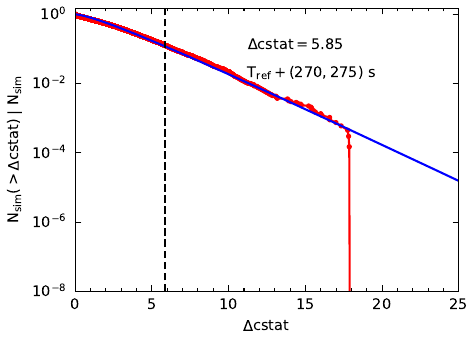}}
 	\vspace{3pt}
 	\centerline{\includegraphics[width=\textwidth]{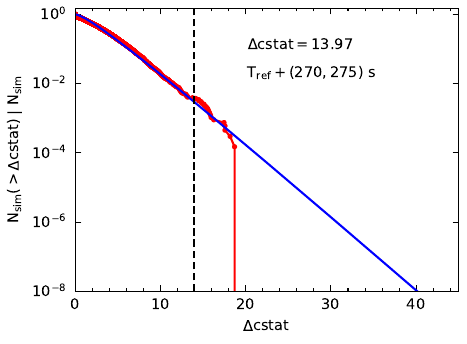}}
 	\vspace{3pt}
 	\centerline{\includegraphics[width=\textwidth]{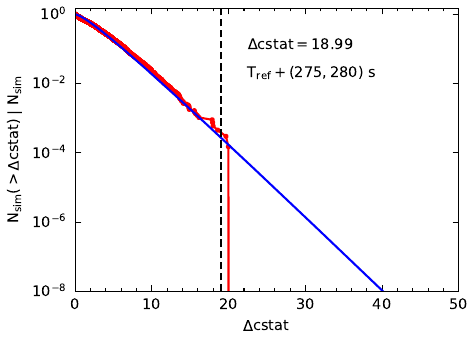}}
 	\vspace{3pt}
 	\centerline{\includegraphics[width=\textwidth]{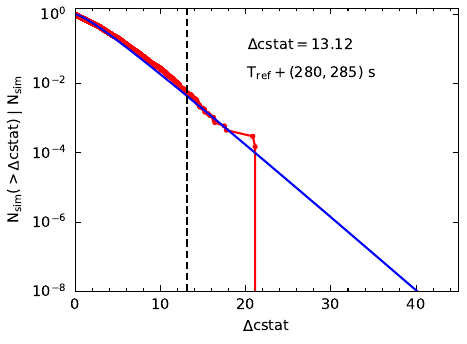}}
 	\vspace{3pt}
 \end{minipage}
 \begin{minipage}{0.32\linewidth}
	\vspace{3pt}
  	\centerline{\includegraphics[width=\textwidth]{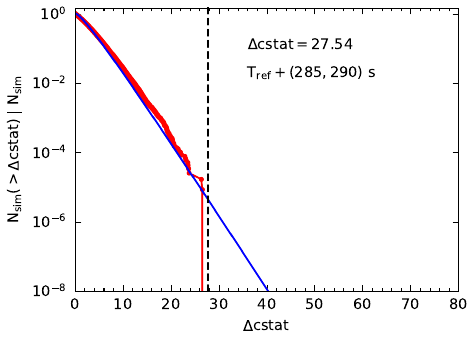}}
 	\vspace{3pt}
	\centerline{\includegraphics[width=\textwidth]{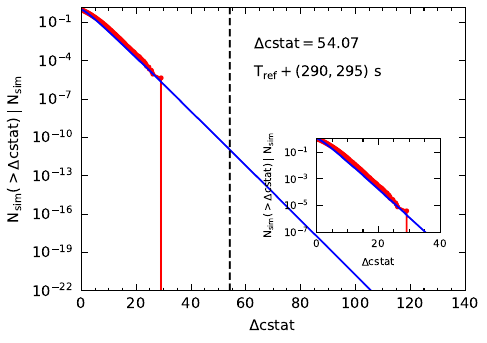}}
	\vspace{3pt}
	\centerline{\includegraphics[width=\textwidth]{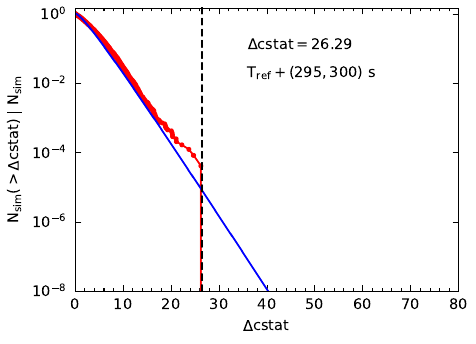}}
	\vspace{3pt}
	\centerline{\includegraphics[width=\textwidth]{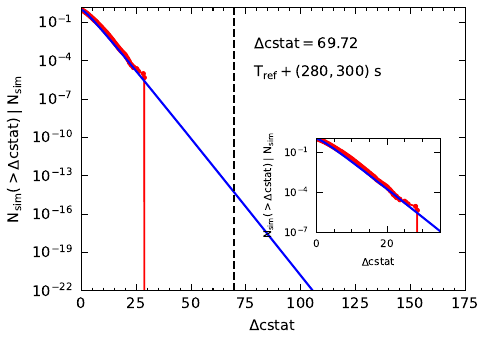}}
	\vspace{3pt}
	\centerline{\includegraphics[width=\textwidth]{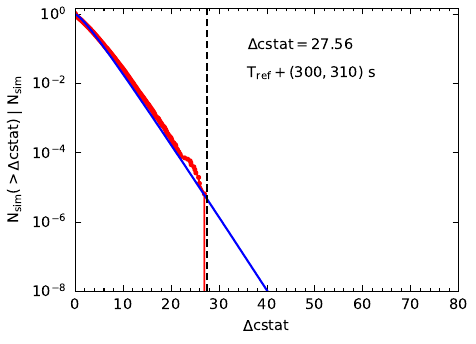}}
	\vspace{3pt}
\end{minipage}
\begin{minipage}{0.32\linewidth}
	\vspace{3pt}
 	\centerline{\includegraphics[width=\textwidth]{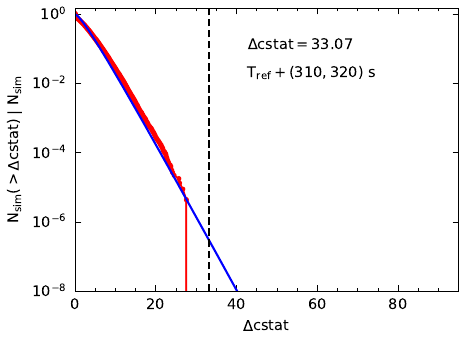}}
	\vspace{3pt}
	\centerline{\includegraphics[width=\textwidth]{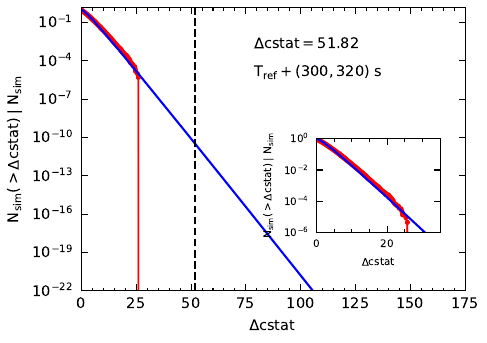}}
	\vspace{3pt}
	\centerline{\includegraphics[width=\textwidth]{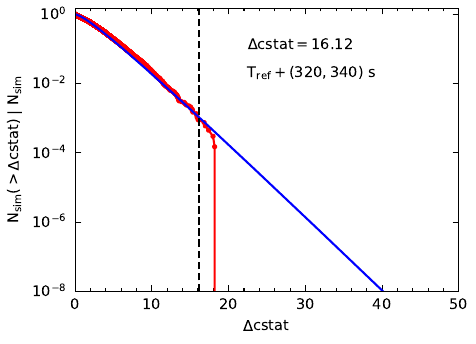}}
	\vspace{3pt}
	\centerline{\includegraphics[width=\textwidth]{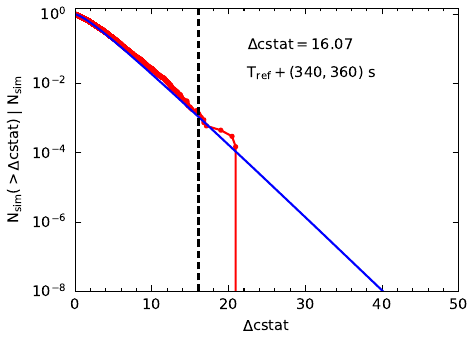}}
	\vspace{3pt}
 	\centerline{\includegraphics[width=\textwidth]{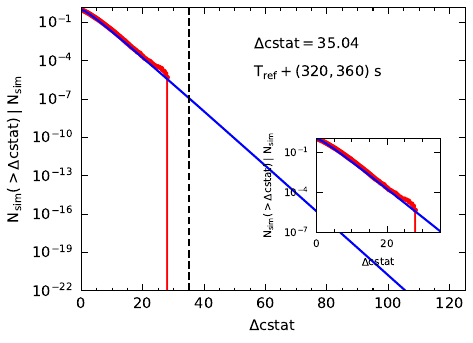}}
	\vspace{3pt}
	% \centerline{Image}
\end{minipage}

\caption{The red data points represent the results from the simulations, while the blue points correspond to the theoretical $\chi^2$ distribution with 3 degrees of freedom. The black dashed line illustrates the difference in the cstat statistic value with and without the gaussian spectral line in the actual fitting result. With the simulations, we can observe that the simulation results follow the theoretical $\chi^2$ distribution. In cases where the simulation covers the specific time periods, we utilize the actual simulation results to calculate the $p$-value. For data points with a large $\rm \Delta cstat$, it would be impractical to calculate the results with simulations. Therefore, we reported the $p$-value as \textless 1/N (where N is the number of simulations). At the same time we also have calculated the theoretical value. }

\label{fig:Significance}
\end{figure*}

As demonstrated in \cref{fig:fitting}, the spectrum fitting results for different time intervals within the time ranges: 246-256 s (G1), 270-275 s (G2), and 275-360 s (G3) are presented. The count spectrum, $\nu F_{\nu}$ and fitted residual plots are displayed for each time interval. It is important to note that different energy ranges and models were utilized for different time periods. We note that all residuals fluctuate around zero, indicating that the model effectively fits the data points.

\subsection{\label{sec:Calculation of The Lorentz Factor} Calculation of the Lorentz factor}
The relationship between the emission line center energy in the jet co-moving coordinate and the observer coordinate is: $\frac{E_{\rm line,co-moving}}{(1+z)} \cdot \Gamma  = E_{\rm line,observer}$, where $\Gamma$ is the (apparent) Lorentz factor of the jet, and $E_{\rm line,co-moving}$ is intrinsic energy of the emission line in the co-moving frame, which is considered to be the electron-positron pair annihilation line (511 keV) in this work, $E_{\rm line,observer}$ is the observed line energy and $z$ is the redshift. From our analysis, we find that $E_{\rm line,observer}=10^m\cdot (t-t_0)^{k}$, thus $\Gamma=\Gamma_0(t-t_0)^{k}$, with which we can estimate the Lorentz factor at any time $t$ after the initial time $t_0$.

\newpage

\onecolumn

\begin{sidewaystable*}
\setlength\tabcolsep{3pt}
\centering
\footnotesize  
\topcaption{Spectral fitting results for GRB 221009A with  GECAM-C and \emph{Fermi}/GBM with spectral line component}

\begin{supertabular*}{\hsize}{@{}@{\extracolsep{\fill}}cccccccccccccc@{}}
\toprule
% \midrule
% \hline 
  Time range & $\alpha$ & $\beta$ & $E_{\rm c}$ (keV) & $\rm PL_{index}$  &LineE (keV) & Sigma (keV)  & factor & cstat/dof &  chi/dof & AIC & $p$-value$_{\rm sim}$ & $p$-value$_{\rm chi}$ & $\rm Flux (\rm erg/cm^2/s$)  \\
   \hline 

(246,256) s & $-1.34 \pm 0.01$ & $-2.86 \pm 0.05$ & $798 \pm 50$ & -  & $37159 \pm 5340$ & $7087 \pm 3101$ & $1.05 \pm 0.08$ & 251/249 & 255/258 & 269 & \textless $4.89\times 10^{-6}$ & $7.17\times 10^{-35}$ & ${2.02}_{-0.06}^{+2.67} \times 10^{-5}$ \\

(270,275) s $^{a}$ & \multirow{2}*{$-1.55 \pm 0.05$} & \multirow{2}*{$-2.76 \pm 0.06$} & \multirow{2}*{$3434 \pm 593$} & -  & $1867 \pm 189$ & $144 \pm 77 $ & \multirow{2}*{$1.12 \pm 0.11$} & \multirow{2}*{207/167} & \multirow{2}*{210/178} & \multirow{2}*{229} & $1.28\times 10^{-1}$ & $1.19\times 10^{-1}$ & ${9.13}_{-5.12}^{+2.59} \times 10^{-6}$ \\

(270,275) s $^{b}$ &  &  & & -  & $17839 \pm 760$ & $1473 \pm 964 $ &  &  &  &  &  $3.90\times 10^{-3}$ & $2.95\times 10^{-3}$ &  ${4.02}_{-1.76}^{+2.04} \times 10^{-6}$\\

(275,280) s & $-1.42 \pm 0.08$ & $-2.69 \pm 0.05$ & $1959 \pm 405$ & -  & $16088 \pm 564$ & $1667 \pm 720$ & $1.18 \pm 0.10$ & 128/127 & 127/135 & 144 &  $4.50\times 10^{-4}$ & $2.74\times 10^{-4}$ & ${3.32}_{-0.72}^{+1.23} \times 10^{-6}$\\
(280,285) s & $-1.49 \pm 0.13$ & $-2.60 \pm 0.10$ & $1740 \pm 637$ & -  & $15016 \pm 536$ & $1995 \pm 677$ & $1.11 \pm 0.14$ & 93/112 & 92/120 & 109 &  $5.55\times 10^{-3}$ & $4.37\times 10^{-3}$ & ${3.98}_{-0.28}^{+3.42} \times 10^{-6}$\\
(285,290) s & $-1.56 \pm 0.01$ & $-2.59 \pm 0.10$ & $1698 \pm 244$ & -  & $13552 \pm 587$ & $1934 \pm 689$ & $1.09 \pm 0.14$ & 240/280 & 237/288 & 256 &  $8.72\times 10^{-6}$ & $4.53\times 10^{-6}$ & ${3.31}_{-0.60}^{+1.55} \times 10^{-6}$\\
(290,295) s & $-1.63 \pm 0.02$ & $-2.44 \pm 0.06$ & $1348 \pm 357$ & -  & $12067 \pm 311$ & $1404 \pm 356$ & $1.35 \pm 0.20$ & 270/313 & 268/321 & 286 &  \textless $4.49\times 10^{-6}$ & $1.09\times 10^{-11}$ & ${2.94}_{-0.51}^{+1.37} \times 10^{-6}$\\
(295,300) s & $-1.67 \pm 0.04$ & $-2.17 \pm 0.06$ & $873 \pm 377$ & -  & $12445 \pm 288$ & $830 \pm 387$ & $1.42 \pm 0.20$ & 272/313 & 269/321 & 288 &  $4.21\times 10^{-5}$ & $8.30\times 10^{-6}$ & ${1.56}_{-0.24}^{+0.72} \times 10^{-6}$\\
(280,300) s & $-1.57 \pm 0.01$ & $-2.41 \pm 0.07$ & $1501 \pm 144$ & -  & $12777 \pm 278$ & $1653 \pm 333$ & $1.09 \pm 0.12$ & 244/237 & 244/245 & 260 &  \textless $4.84\times 10^{-6}$ & $4.91\times 10^{-15}$ & ${2.62}_{-0.33}^{+0.53} \times 10^{-6}$\\

(300,310) s & - & - & - & $-1.92 \pm 0.01$ & $10150 \pm 328$ & $983 \pm 382$ & $0.82 \pm 0.06$ & 273/315 & 270/321 & 285 & $6.64\times 10^{-6}$ & $4.49\times 10^{-6}$ & ${1.74}_{-0.47}^{+0.51} \times 10^{-6}$\\
(310,320) s & - & - & - & $-2.02 \pm 0.02$ & $8871 \pm 489$ & $1740 \pm 529$ & $1.16 \pm 0.13$ & 279/315 & 278/321 & 291 &  \textless $4.55\times 10^{-6}$ & $3.11\times 10^{-7}$ & ${1.58}_{-0.31}^{+1.02} \times 10^{-6}$\\
(300,320) s & - & - & - & $-1.96 \pm 0.01$ & $9726 \pm 256$ & $1115 \pm 290$ & $0.96 \pm 0.06$ & 264/315 & 263/321 & 276 & \textless $4.93\times 10^{-6}$ & $3.27\times 10^{-11}$ & ${1.51}_{-0.25}^{+0.39} \times 10^{-6}$\\

(320,340) s & $-1.33 \pm 0.56$ & - & $89 \pm 57$ & $-2.00 \pm 0.12$ & $6097 \pm 2361$ & $3441 \pm 1779$ & $1.00-$ & 276/313 & 274/321 & 292 &  $9.00\times 10^{-4}$ & $1.07\times 10^{-3}$ & ${1.40}_{-0.59}^{+0.04} \times 10^{-6}$\\
(340,360) s & $-1.59 \pm 0.11$ & - & $116 \pm 26$ & $-1.88 \pm 0.12$ & $6374 \pm 622$ & $2342 \pm 575$ & $1.00-$ & 304/313 & 303/321 & 320 &  $1.50\times 10^{-3}$ & $1.10\times 10^{-3}$ & ${1.60}_{-0.12}^{+0.54} \times 10^{-6}$\\
(320,360) s & $-1.43 \pm 0.22$ & - & $93 \pm 28$ & $-1.97 \pm 0.08$ & $6310 \pm 861$ & $2871 \pm 707$ & $0.98 \pm 0.17$ & 296/325 & 295/334 & 314 & \textless $4.98\times 10^{-6}$ & $1.20\times 10^{-7}$ & ${1.39}_{-0.24}^{+0.52} \times 10^{-6}$\\

\hline  
\end{supertabular*} \label{Spectrum fitting results with Gauss Component}

\footnotesize{{Flux is calculated in the gaussian 3$\rm \sigma$ energy range; In the time period \T+(270,275) s, we find the existence of a gaussian absorption line and a gaussian emission line, which we mark in the table with a,b respectively. $p$-value$_{\rm sim}$ is calculated based on the actual simulation results, and if a sufficient number of simulations cannot be performed, we report \textless 1/N. $p$-value$_{\rm chi}$ is calculated theoretically based on the likelihood ratio approximately following the chi-square distribution.}}\\

\end{sidewaystable*}

\newpage

\begin{sidewaystable*}
\centering
% \small 
\topcaption{Spectral fitting results for GRB 221009A with GECAM-C and \emph{Fermi}/GBM  without spectral line component}
\begin{supertabular*}{\hsize}{@{}@{\extracolsep{\fill}}cccccccccc@{}}
\toprule
% \hline 
  Time range & $\alpha$ & $\beta$ & $E_{\rm c}$ (keV) & $\rm PL_{index}$ & factor & cstat/dof &  chi/dof & AIC & $\rm \Delta AIC $ \\
   \hline 
   
(246,256) s & $-1.34 \pm 0.01$ & $-2.63 \pm 0.03$ & $750 \pm 48$ & -  & $0.79 \pm 0.04$ & 413/252 & 408/258 & 425 &  156\\
   
(270,275) s & $-1.57 \pm 0.04$ & $-2.66 \pm 0.03$ & $3292 \pm 572$ & -  & $1.18 \pm 0.09$ & 224/173 & 231/178 & 234 &  5 \\

(275,280) s & $-1.42 \pm 0.08$ & $-2.65 \pm 0.05$ & $1952 \pm 415$ & -  & $1.23 \pm 0.11$ & 146/130 & 146/135 & 156 &  12\\
(280,285) s & $-1.50 \pm 0.16$ & $-2.35 \pm 0.07$ & $1680 \pm 861$ & -  & $0.88 \pm 0.11$ & 122/115 & 121/120 & 132 &  23\\
(285,290) s & $-1.56 \pm 0.02$ & $-2.35 \pm 0.06$ & $1523 \pm 252$ & -  & $0.95 \pm 0.10$ & 265/283 & 262/288 & 275 &  19\\
(290,295) s & $-1.61 \pm 0.03$ & $-2.25 \pm 0.04$ & $947 \pm 273$  & -  & $1.37 \pm 0.15$ & 322/316 & 316/321 & 332 &  46 \\
(295,300) s & $-1.66 \pm 0.05$ & $-2.06 \pm 0.05$ & $752 \pm 330$  & -  & $1.23 \pm 0.16$ & 296/316 & 288/321 & 306 &  18 \\
(280,300) s & $-1.53 \pm 0.01$ & $-2.11 \pm 0.04$ & $900 \pm 113$  & -  & $0.78 \pm 0.07$ & 314/240 & 312/245 & 324 &  64 \\

(300,310) s & - & - & - & $-1.91 \pm 0.01$ & $0.88 \pm 0.06$ & 298/318 & 292/321 & 304 &  19\\
(310,320) s & - & - & - & $-2.01 \pm 0.02$ & $1.34 \pm 0.12$ & 311/318 & 307/321 & 317 &  26\\
(300,320) s & - & - & - & $-1.95 \pm 0.01$ & $1.05 \pm 0.06$ & 315/318 & 311/321 & 321 &  45\\

(320,340) s & $-1.59 \pm 0.12$ & - & $115 \pm 31$ & $-1.83 \pm 0.07$ & $1.00-$ & 290/316 & 288/321 & 300 &  8\\
(340,360) s & $-1.57 \pm 0.09$ & - & $101 \pm 19$ & $-1.79 \pm 0.06$ & $1.00-$ & 319/316 & 317/321 & 329 &  -9\\
(320,360) s & $-1.54 \pm 0.12$ & - & $101 \pm 23$ & $-1.86 \pm 0.06$ & $1.12 \pm 0.24$ & 330/328 & 327/334 & 342 &  28\\
\hline 
\end{supertabular*} \label{Spectrum fitting results without Gauss Component}
\end{sidewaystable*}

%\end{appendices}
\end{appendix}

\end{multicols}
\end{document}